\documentclass[review]{elsarticle}

\usepackage{lineno,hyperref}
\modulolinenumbers[5]

\journal{Scandinavian Journal of Statistics}

\usepackage{epsfig}
\usepackage{float}
\usepackage{amsmath,amsthm,verbatim,amssymb}
\usepackage{verbatim,color,amsmath}
\usepackage{lscape}
\usepackage{mathrsfs}

\renewcommand{\baselinestretch} {1.3}
\makeatletter 
\def\singlespace{\def\baselinestretch{1}\@normalsize}

\date{}

\newtheorem{theorem}{Theorem}
\newtheorem{lemma}{Lemma}

\def\bm{\bf}
\def\ihe{{\sum_{i=1}^{n}}}

\def\l{\left}
\def\r{\right}

\newcommand{\bZ}{{\bf Z}}

\newcommand{\bz}{\mbox{\bf z}}
\newcommand{\ba}{\mbox{\bf a}}

\newcommand{\bB}{\mbox{\bf B}}
\newcommand{\bA}{\mbox{\bf A}}

\newcommand{\bb}{\mbox{\bf b}}
\newcommand{\bab}{\mbox{\bf ab}}

\newcommand{\bV}{\mbox{\bf V}}

\newcommand{\bbeta}{\boldsymbol{\beta}}

\newcommand{\bSigma}{\boldsymbol{\Sigma}}

\newcommand{\btheta}{\boldsymbol{\theta}}

\newcommand{\rmd}{\rm d}

\newcommand{\bse}{\begin{eqnarray*}}
\newcommand{\ese}{\end{eqnarray*}}

\newcommand{\rmT}{\rm T}

\def\wh{\widehat}
\def\wt{\widetilde}
\def\overline{\bar}

\begin{document}

\begin{frontmatter}

\title{Cox regression analysis for distorted covariates with an unknown distortion function}

\author[a]{Yanyan Liu}
\author[b]{Yuanshan Wu}
\author[b]{Jing Zhang\corref{cor}}
\author[c]{Haibo Zhou\corref{cor}}

\address[a]{School of Mathematics and Statistics, Wuhan University, Wuhan, Hubei 430072, China}
\address[b]{School of Statistics and Mathematics, Zhongnan University of Economics and Law, Wuhan, Hubei 430073, China}
\address[c]{Department of Biostatistics, University of North Carolina at Chapel Hill, Chapel Hill, NC
27599-7420, USA}

\cortext[cor]
{Corresponding author: jing66@zuel.edu.cn(Jing Zhang), zhou@bios.unc.edu(Haibo Zhou)}

\begin{abstract}
We study inference for censored
survival data where some covariates are distorted by some unknown
functions of an observable confounding variable in a multiplicative
form.
Example of this kind of data in medical studies is the common practice to
normalizing some important observed exposure variables by patients'
body mass index (BMI), weight or age.
Such phenomenon also appears frequently in environmental studies where
ambient measure is used for normalization,
and in genomic studies where library size needs to be normalized for
next generation sequencing data.
We propose a new covariate-adjusted Cox proportional hazards regression model and
utilize the kernel smoothing method to estimate the distorting function,
then employ an estimated maximum likelihood method to derive
estimator for the regression parameters.
We  establish the large sample properties of the proposed estimator.
Extensive simulation studies
demonstrate that the proposed estimator performs well
in correcting the bias arising from distortion.
A real data set from  the National Wilms' Tumor Study (NWTS)
is used to illustrate the proposed approach.
\end{abstract}

\end{frontmatter}


\section{Introduction}
In real studies, the primary covariates sometimes
are not directly recorded in their true values,
but rather, they are observed in a distorted form,
where the distortion is in the form of
a multiplicative factor.
This type of data does not get sufficient attention as other
types of covariate measurement error problems,
even though they are
also quite wide prevalent in real studies.
For example,
when  releasing household data on energy use,
in order to maintain confidentiality,
the U.S. Department of Energy  multiplied the survey data
by some randomly selected numbers before publication (Hwang, 1986).
Therefore,  the contaminated data available to
the public is $\widetilde{X}=X\cdot U$,
where $X$ and $U$ respectively denote the true data
and the randomly selected number.
This multiplicative contamination structure is also very common in
biomedical studies,
in the form of normalization,
as some primary covariates are often
normalised by a confounder such as
BMI ($\textrm{BMI}=\textrm{weight}/\textrm{height}^2$)
or by other measures of body configuration or age.
For instance,
in a study of the relationship between the fibrinogen level (FIB)
and serum transferrin level (TRF) among hemodialysis patients,
Kaysen et al. (2002) found that BMI
has a great influence on FIB and TRF and may distort the true
relationship between them.
Therefore, they proposed a calibration method where
dividing the observed FIB and TRF  by the confounding variable BMI.
This implies a multiplicative structure
between the unobserved primary variables and the confounding variable.
Such phenomenon also appears frequently in environmental studies where
ambient measure is used for normalization,
and in genomic studies where library size needs to be normalized for
next generation sequencing data.

In some situations, however,
the precise nature of multiplicative relationship between
the primary variables and the confounding variable
could be  unknown,
and in this case
the naive practice of dividing  by the confounding variable
may result in biased estimates or losing of power for statistical inference.
To overcome these difficulties,
Sent$\ddot{\mbox{u}}$rk \& M$\ddot{\mbox{u}}$ller (2005)
considered a more flexible multiplicative form
which is  an unknown function
of the confounding variable $U$.
They  proposed a
covariate adjustment method for the linear regression model,
where  both the response $(Y)$ and the covariates $(X)$ are distorted by
an observable confounder $U$, i.e.,
$\wt{X}=\phi(U)X$, $\wt{Y}= \varphi(U)Y$,
where $ \wt{X}$ and $\wt{Y}$  are observable distorted covariates and response,
$\phi(\cdot)$ and $\varphi(\cdot)$ are unknown smooth distorting functions.
Directly applying the widely used ordinary least squares (OLS) method to the
contaminated data $(\wt{X}, \wt{Y})$ will result in biased and inconsistent estimates.
Sent{\"u}rk \& M{\"u}ller (2005)  corrected the bias by
linking  it to a varying-coefficient regression model,
then utilized the bin method (Fan \& Zhang, 2000) to obtain consistent estimators (Sent{\"u}rk \& M{\"u}ller, 2006).
Related research includes
Nguyen \& Sent{\"u}rk (2008) on generalizing
this method  to the case of multiple distorting covariates,
Sent{\"u}rk \& M{\"u}ller (2009) on extending  to generalized linear model,
Zhang, Zhu, \& Liang (2012) and Zhang et al. (2013) on the nonlinear regression model and  the partial linear model.
More recently, Cui et al. (2009) developed a direct plug-in
estimation procedure for nonlinear regression model with one confounding variable.
They  proposed to estimate
the distorting functions $\varphi(\cdot)$ and $\phi(\cdot)$
by nonparametrically regressing the response and predictors
on the distorting variable,
and obtained the estimates $(\wh{X}, \wh{Y})$
for the unobservable response and predictors,
then conducted the nonlinear least squares method on
the estimated counterparts $(\wh{X}, \wh{Y})$.
Zhang, Zhu, \&  Zhu (2012) further applied this direct plug-in method
to  semiparametric model by incorporating dimension reduction techniques.
To relax the parametric assumptions and some
restrictive conditions on distorting functions in the existing literature,
Delaigle, Hall, \& Zhou (2016) proposed a more flexible
nonparametric estimator for the regression function.

In this paper,  we focus on investigating censored
survival data where the response of interest is a right-censored survival time
and the primary predictor $X$ is
distorted by an observable confounding variable $U$ through
the multiplicative form $\wt{X}=\phi(U)X$,
where $\phi(\cdot)$ is the unknown distorted function.
A reasonable identifiability condition for this structure is
$E\{\phi(U)\}=1$ corresponding to the assumption that
the mean distorting effect vanishes (Sent{\"u}rk \& M{\"u}ller, 2005).
The existing methods mentioned earlier can not be applicable here
due to censoring.
Furthermore, the existing methods for
censored survival data with mismeasured covariates
(e.g., Prentice, 1982; Wang et al., 1997;
Zhou \& Pepe, 1995; Zhou \& Wang, 2000;
Huang \& Wang, 2000; Hu \& Lin, 2002)
can not handle this multiplicative distortion.
To make valid inference,
we propose a covariate-adjusted Cox proportional
hazards regression to address this multiplicative contamination structure.
Inspired by Cui et al. (2009), we first
employ the nonparametric regression
to obtain consistent estimator of
the distorting function $\phi(\cdot)$ through the kernel smoothing method,
and obtain the estimates for the true covariates $X$
by $\hat{X}=\tilde{X}/{\hat{\phi}(U)}$.
Then the regression parameters are estimated by maximizing
the partial likelihood on the estimated data.
Our approach has several distinctive advantages.
First,
the contamination structure we considered
is more general which includes a large class of confounding mechanisms, e.g.,
$\phi(\cdot)=1$ means there is no contamination,
$\phi(U)=U$ represents the contamination structure $\wt{X} = X\cdot U$.
So the applicability of our proposed method can be quite broad.
Second, the computation of our method is simple and fast, which will greatly facilitates
its implementation in real application.

The rest of the article is organized as follows.
In Section $2$, we introduce the covariate-adjust Cox regression
for the  multiplicative contaminated data
and present the proposed covariate-calibration method.
In Section $3$,
we establish the asymptotic properties of the proposed estimates.
In Section $4$, we present
simulation results to evaluate the
finite sample performance of the proposed estimates.
In Section $5$, we apply the
proposed method to a data set from the National Wilms' Tumor Study (NWTS).
Some concluding remarks are given in Section $6$.
All technical proofs are presented in the supplementary material.

\section{Cox regression with multiplicative contamination structure}
\subsection{Model, data and contamination structure}
To fix notation,
let $T$ denote the survival time, $C$ denote the censoring time,
${\wt{T}}=\min({T},C)$ denote the observed time,
and $\Delta=I(T\leq C)$ denote the failure indicator.
Let ${\bf Z}=(Z_{1},Z_{2},\ldots,Z_{p})^{\rm T}$ and $X$
be the associated covariates where $X$ is the one that subjects to
multiplicative contamination.
Assume that the censoring mechanism is random,
that is, the survival time $T$ and
the censoring time $C$ are conditionally
independent given $\bf Z$ and $X$.
The proportional hazards  regression model (Cox, 1972) assumes
that the conditional hazard function of the survival
time ${T}$ associated  with covariates $\bf Z$
and $X$ takes the form of
\begin{eqnarray*}
\lambda(t|{\bf Z},X)=\lambda_0(t)\exp({\bbeta}^{\rm {T}} {\bf Z}+{{\gamma}} X),
\end{eqnarray*}
where $\lambda_0(t)$ is the baseline hazard function,
${\bbeta}=(\beta_{1},\beta_{2},\ldots,\beta_{p})^{\rm T}$  and $\gamma$
are the unknown regression coefficients.
We assume the scalar covariate $X$ is not
observed precisely while the $p$-dimensional covariate $\bZ$
could be accurately observed.
Assume the observed data consists of $n$ subjects, denoted by
$(\wt{T}_i,\Delta_i,{\bf Z}_i,\\U_i,\wt{X}_i)$, $i=1,\ldots,n,$
which are independent samples from $(\wt{T},\Delta,{\bf Z},U,\wt{X})$.
Instead of  exact $X_i$, we observe
$\wt{X}_i$ such that
\begin{eqnarray}\label{errstruct}
\wt{X}_i=\phi(U_i)X_i,
\end{eqnarray}
where $U_i$ is an observable variable and independent of $X_i$,
$\phi(\cdot)$ is an unknown link function.
To make the model identifiability,
we assume that $E\{\phi(U_i)\}=1$, which
implies that the distorting effect vanishes on average.

We aim to infer the regression parameters
$\bbeta$ and $\gamma$ based on the observations available.
When $X_i$ are observed without contamination,
maximizing the partial likelihood (Cox, 1975)
\begin{eqnarray}\label{pl}
L_n({\bbeta},\gamma)=\prod\limits_{i=1}^n\left\{\frac{\exp(\bbeta^{\rmT}\bZ_i+\gamma X_i)}
{\sum_{j=1}^nI(\wt{T}_j>\wt{T}_i)\exp(\bbeta^{\rmT}\bZ_j+\gamma X_j)}\right\}^{\Delta_i}
\end{eqnarray}
can offer the estimates for $\bbeta$ and $\gamma$.
It is evident that (\ref{pl}) can not be used
when $X_i$ are unobservable or contaminated.

Note that the established methods on the Cox regression
with additive contamination structure
$\wt{X} = X+U$
always require  error $U$ to be independent of
$X$ (e.g., Huang \& Wang, 2000; Li \& Ryan, 2004).
Directly apply the additive error structure methods
to current setting is not feasible.
To illustrate this, even though
the multiplicative contamination structure (\ref{errstruct}) can
also be rewritten as
an additive structure,
\begin{equation}
\label{eq1}\wt{X}=X+X\{\phi(U)-1\},
\end{equation}
or
\begin{equation}
\label{eq2}\log \wt{X}=\log X+\log\{\phi(U)\},
\end{equation}
the  error $X\{\phi(U)-1\}$ is not independent of $X$,
hence the methods mentioned above can not be applicable here.
If one takes the logarithmic
transformation assuming the related quantities
are positive, then one would arrive at the additive
covariate contamination  structure (\ref{eq2}).
Here the error term $\log\{\phi(U)\}$ is independent of $\log X$,
but extra variation needs to be accounted
for in the back-transformation procedure.
Moreover, the routine approximately corrected
score method for the Cox regression
at the scale $\log X$ would result in biased
estimate if the correct Cox regression model is
 linear in $X$.

\subsection{Covariate-calibration method}
Our proposed approach is based on
directly calibrating $X_i$.
Note that
\begin{eqnarray*}
\phi(u)=\frac{E(\wt{X}|U=u)}{E (X)}=\frac{E(\wt{X}|U=u)}{E (\wt{X})}.
\end{eqnarray*}
We can employ the commonly used Nadaraya--Watson kernel smoothing estimate for
$\psi(u)=E(\wt{X}|U=u)$, which is given by
\begin{eqnarray*}
\wh\psi(u)=\frac{\sum_{i=1}^n K\{(u-U_i)/h_n\}\wt{X}_i}{\sum_{i=1}^n K\{(u-U_i)/h_n\}},
\end{eqnarray*}
where $K(\cdot)$ is the kernel smoothing function
and $h_n$ is the bandwidth.
Since $\bar{\wt{X}}_n=n^{-1}\sum_{i=1}^n\wt{X}_i$ converges to $E(\wt{X})$ almost surely
by using the  strong law of large numbers,
we can obtain an consistent estimate for $\phi(u)$ as
$\wh{\phi}(u)=\wh\psi(u)/ \bar{\wt{X}}_n$.
Following (\ref{errstruct}),  we propose a calibration
of $X_i$  by
$\wh{X}_i={\wt{X}_i}/{\wh{\phi}(U_i)}$.
 Therefore, we can construct an estimated
partial likelihood using $\wh{X}_i$ as follows,
\begin{eqnarray}\label{epl}
\wh L_n({\bbeta},\gamma)=\prod\limits_{i=1}^n\left\{\frac{\exp(\bbeta^{\rmT}\bZ_i+\gamma \wh X_i)}
{\sum_{j=1}^nI(\wt{T}_j>\wt{T}_i)\exp(\bbeta^{\rmT}\bZ_j+\gamma \wh X_j)}\right\}^{\Delta_i}.
\end{eqnarray}
The proposed estimator
$(\wh\bbeta,\wh\gamma)$ was defined as the maximizer for $\wh L_n({\bbeta},\gamma)$,
i.e.,
\begin{eqnarray}\label{epll}
(\wh\bbeta,\wh\gamma)={\rm argmax}_{(\bbeta,\gamma)}\wh L_n({\bbeta},\gamma).
\end{eqnarray}

\subsection{Bandwidth selection}
In real data analysis, it is desirable to have an automatically data-driven method
for selecting the bandwidth parameter $h_n$.
We will employ a cross-validation (CV) method to choose the optimal $h_n.$
The kernel estimate of the  density function of $U$,  $p(u)$, is denoted as
\begin{eqnarray*}
  \wh{p}(u)=\frac{1}{nh_n}\sum_{i=1}^n K\left(\frac{u-U_i}{h_n}\right).
\end{eqnarray*}
Following Rudemo (1982) and Bowman (1984), we define an integrated
squared error (ISE) as follows,
\begin{eqnarray}\label{ISE}
{\rm ISE}(h_n)&=&\int\left\{\wh{p}(u)-p(u)\right\}^2 {\rm d}u\nonumber
\\
&=&\int\left\{\wh{p}(u)\right\}^2 {\rm d}u-2\int\wh{p}(u)p(u) {\rm d}u+
\int \left\{p(u)\right\}^2 {\rm d}u.
\end{eqnarray}
As the third term of (\ref{ISE}) is free of
$h_n$, the minimizer of the ISE$(h_n)$ is the same as the minimizer
of the sum of the first two terms of (\ref{ISE}).
Let $\wh{p}_{(-i)}(\cdot)$
be the leave-one-out kernel density estimator, i.e.,
\begin{eqnarray*}
  \wh{p}_{(-i)}(u)=\frac{1}{nh_n}\sum_{j\neq i}^n K\left(\frac{u-U_j}{h_n}\right).
\end{eqnarray*}
The second term of (\ref{ISE}) can be consistently
estimated by $-2n^{-1}\sum_{i=1}^n \wh{p}_{(-i)}(U_i).$
Therefore, we propose a cross-validation criterion as follows,
\begin{eqnarray*}
 {\rm CV}(h_n)=\int\left \{\wh{p}(u)\right\}^2 {\rm d}u-2n^{-1}\sum_{i=1}^n \wh{p}_{(-i)}(U_i).
\end{eqnarray*}
Denote
\begin{eqnarray*}
\wh{h}_{n,{\rm opt}}={\rm argmin}_{h_n}{\rm CV}(h_n),
\end{eqnarray*}
which is considered as the optimal bandwidth parameter.

\section{Asymptotic properties}
We set $\btheta = (\bbeta^{\rm T}, \gamma)^{\rm T}$,
let $\wh{\btheta}=(\wh{\bbeta}^{\rm T},\wh{\gamma})^{\rm T}$
and ${\btheta}_0=({\bbeta}_0^{\rm T},\gamma_0)^{\rmT}$ respectively represent
the estimation and the true value of the regression parameter
$\btheta$.
The following theorem gives the consistency and asymptotic normality
of the proposed estimator $\wh{\btheta}$  when $n\rightarrow \infty$.
The regularity conditions and the proofs of this theorem are given in the Appendixes A and
B, respectively.

 \begin{theorem}
 Let $\wh{\btheta}=(\wh{\bbeta}^{\rm T},\wh{\gamma})^{\rm T}$ be defined by (\ref{epll}).
If conditions C1-C9 in the Appendix A are satisfied, the following results hold:
\begin{itemize}
  \item[(\romannumeral 1)]
   $\wh{\btheta}$ coverges in probability to the true value ${\btheta}_0$,
  \item[(\romannumeral 2)]
  $\sqrt n (\wh{\btheta}-{\btheta}_0)\stackrel{d}{\longrightarrow} \text{N}\l({\bf0},\Sigma^{-1}(\Sigma+\Omega)\Sigma^{-1}\r)$,
  \end{itemize}
where
  $\Sigma=
  \begin{pmatrix}
  \Sigma_{11} & \Sigma_{12}\\
  \Sigma_{21} & \Sigma_{22}\\
 \end{pmatrix}$
 is defined in condition C4, $\Sigma_{11}$ denotes the $p$th order sequential principal minor
 of $\Sigma$,
  $\zeta=(-\Sigma_{12}^{\rm T} \gamma_0, -\Sigma_{22} \gamma_0)^{\rm T}$,
  $\Omega=\frac{\text{Var}(\wt{X})-\text{Var}(X)}{\l\{E(X)\r\}^2} \zeta \zeta^{\rm T}$.

 \end{theorem}

The above theorem establishes the asymptotic normality of the proposed estimator
$\wh{\btheta}$, furthermore,
from (\romannumeral 2) in Theorem 1, we can
obtain the asymptotic distribution of
$\wh{\bbeta}$ and $\wh{\gamma}$ respectively, i.e.,
$\sqrt n (\wh{\bbeta}-{\bbeta}_0)\stackrel{d}{\longrightarrow} \text{N}\l({\bf0},(\Sigma^{-1})_{p}\r)$,
and
$\sqrt n (\wh{\gamma}-{\gamma}_0)\stackrel{d}{\longrightarrow}\text{N}\l({0},(\Sigma^{-1})_{(p+1,p+1)}+\frac{\text{Var}(\wt{X})-\text{Var}(X)}{\l\{E(X)\r\}^2}\gamma_0^2\r)
$, where
$(\Sigma^{-1})_{p}$ and $(\Sigma^{-1})_{(p+1,p+1)}$
respectively denote the $p$th order sequential principal minor
and the $(p+1)$th diagonal element of matrix $\Sigma^{-1}$.
We note a few remarks on the terms in the expression
for the asymptotic covariance matrix.
If there is no distortion
with $\phi(\cdot)=1$, we can estimate $\btheta$ by
maximizing the partial likelihood  (\ref{pl}),
the asymptotic covariance matrix of
$\wh{\btheta}$ is $\Sigma^{-1}$.
So the term $\Sigma^{-1}\Omega\Sigma^{-1}$
is caused by the distortion.
Furthermore,
the limiting variance for $\wh{\gamma}$
includes some unknown
components to be estimated, therefore,
we can use the sandwich method and plug-in estimation
to obtain the standard error and
construct the confidence region
for $\wh{\gamma}$.

\section{Simulation studies}
We conducted extensive simulations to investigate
the finite-sample performance of the proposed estimator
$(\wh\bbeta_P,\wh\gamma_P)$ and compared it
with two completing estimators.
The first one is the naive estimator $(\wh\bbeta_N,\wh\gamma_N)$ that
ignores the contamination and directly uses $\wt{X}$ to replace $X$;
the second one is the oracle estimate $(\wh\bbeta_O,\wh\gamma_O)$,
which is obtained by assuming that $X$ was known.

The survival times $T_i$ were generated from the Cox proportional hazards model with the
conditional hazard function given by
\vspace{-1ex}
\begin{eqnarray*}
   \lambda(t|{\bZ}_i,X_i)=\lambda_0(t)\exp({\bbeta_0}^{\rm T}{\bZ}_i+\gamma_0 X_i).
\end{eqnarray*}
Set $\bbeta_0=(1,0.5)^{\rm T}$, $\gamma_0=1.5$ and
the baseline hazard function $\lambda_{0}(t)=1$.
The covariate $\bZ_i=(Z_{i1},Z_{i2})^{\rm T}$ follows
a multivariate normal distribution with
 mean $\bf 0$ and correlation matrix
 $\bSigma=(0.8^{|j-k|})$ for $j,k=1, 2$. We generated $X_i$ from $N(1,0.5^2)$ and
the confounding covariates $U_i$ from a uniform distribution over interval $[2,6]$.
We considered two forms of  distortion function $\phi(u)=(u+3)/7$ and
$\phi(u)=3(u+1)^2/79$, which satisfy $E\{\phi(U_i)\}=1$.
We took the censoring time $C=\widetilde{C}\wedge \tau$, where $\widetilde C$
was generated from Unif$(0,\tau+2)$. The study duration $\tau$
was chosen to yield the desirable censoring rate.
To estimate the distorting function, we chose Gaussian kernel function
$K(t)=\exp(-{t^2}/{2})/{\sqrt{2\pi}}$
and adopted
the leave-one-out cross-validation method to select the bandwidth.
We took the sample size $n=100$ and $n=200$, coupled with the censoring rates (CR) of
$20\%$, $40\%$ and $80\%$.
For each configuration, we repeated $1000$ simulations.

Tables 1  and 2 summarize the results of $(\wh\bbeta_P,\wh\gamma_P)$, $(\wh\bbeta_N,\wh\gamma_N)$
and $(\wh\bbeta_O,\wh\gamma_O)$ under
different distortion functions and different censoring rates
for sample size $n=100$ and  $n=200$, respectively. We make the following
observations:
(i) As expected, in terms of the mean-square error or the coverage probability,
the oracle estimator $(\wh\bbeta_O,\wh\gamma_O)$
and our proposed estimator $(\wh\bbeta_P,\wh\gamma_P)$ are
all superior to the naive estimator $(\wh\bbeta_N,\wh\gamma_N)$,
especially for the results of $\wh{\gamma}$. Not surprisingly,
the  naive estimator $(\wh\bbeta_N,\wh\gamma_N)$  are seriously biased.
For example,
under the censoring rate of $20\%$ and $\phi(u)=3(u+1)^2/79$ in Table 1,
the bias for  $\wh\gamma_N$ is $-0.810$, more than half of its real value $1.5$,
while the bias for proposed estimator $\wh\gamma_P$ is only $-0.047$;
moreover,
the coverage probability for $\wh\gamma_N$ is $0.006$, almost equals to zero.
(ii) The proposed estimator $(\wh\bbeta_P,\wh\gamma_P)$
are essentially unbiased and comparable with the oracle estimator
under different settings, even for the cases with high censoring rate of $80\%$.
For example,
in  the case of  censoring rate$=40\%$ and $\phi(u)=3(u+1)^2/79$ in Table 1,
the relative efficiency
$\frac{SD(\wh\gamma_P)}{SD(\wh\gamma_O)}=\frac{0.341}{0.324}=1.05$,
very close to 1.
(iii) Our proposed method performs stably with the choice of the distortion function,
while the naive method performs worse if we chose $\phi(u)=3(u+1)^2/79$.
The coverage probabilities of $\wh\gamma_N$ for $\phi(u)=3(u+1)^2/79$
almost equal or close to zero.
These simulation results demonstrate that
the proposed covariate-calibration approach
can effectively overcome the negative effect arising from
the covariate contamination and meanwhile exhibits good performance.

\section{ Analysis with Wilms' tumor study}
We applied the proposed covariate-calibration method
to the Wilms' tumor data, which was collected in two randomized studies
in Wilms' tumor patients.
Wilms' tumor is a rare kidney cancer occurring in young children.
The National Wilms' Tumor Study Group (NWTSG) conducted
several randomized studies to test different treatments in
Wilms' tumor patients. We use a Wilms' tumor data including $3915$ patients participating
in two of the NWTSG trials NWTS-3 and NWTS-4 (D'Angio et al., 1989; Green
et al., 1998) to evaluate the joint effect of tumor weight, histological type and other risk factors.
The primary endpoint of the study was the survival time (in years).
%
During the follow-up, $444$ patients died
of Wilms' tumor and the other $3471$ patients were censored,
which led to the censoring rate of $88.66\%$.
The mean observed  time was 10.33 years (ranging from 0.01 to 22.50 years).
We divided the data into two groups according to the histological type
(favorable and unfavorable) and summarized
the size and mean of each covariate  in Table 3.
It can be seen that $3476$ patients have favorable tumor and the other $439$ patients have unfavorable tumor.
The mean observed time for patients with favorable tumor  is 10.68 years, which is
larger then the corresponding value (7.55) of the unfavorable tumor group.
Figure 1 shows the Kaplan-Meier curves for the two different tumor histological types,
from which we can  see that patients with favorable tumor experienced longer survival time.

The predictors included in this analysis are the weight of tumor bearing specimen (abbreviated as wgt, in kilograms),
the histological type of the tumor (type, being 0 if favorable and 1 otherwise),
tumor stage (stage, coded by 1 and 0, indicating spread of
the tumor from localized to metastatic),
age at diagnosis (age, measured in years),
the study number (num, 1 denotes NWTS-3 and 0 denotes NWTS-4). 

We examine the following  Cox proportional hazards regression model,
\begin{eqnarray*}
\lambda(t)=\lambda_0(t)\exp(\gamma \cdot \rm{wgt}+\beta_1 \cdot \rm{type}+\beta_2 \cdot \rm{stage}+
  \beta_3 \cdot \rm{age}+\beta_4 \cdot \rm{num}).
\end{eqnarray*}
It is known that   the weight of tumor bearing specimen (wgt) is affected by tumor's  diameter (diam, in centimeters).
The scatter points of wgt versus  diam shown in Figure 2 clearly demonstrate that there indeed exists a strong positive correlation between them.
Therefore, we directly adjust for  the potential distorting covariate with the proposed method and assume the distortion model as
$\wt{\rm{wgt}}=\phi(\rm{diam}) \cdot \rm{wgt}$,
where $\phi(\cdot)$ is an unknown link function, $\wt{\rm{wgt}}$ is the observed wgt.
The analysis results of the covariate effects were summarized in Table 4.
As a comparison, we also presented the results of the naive method which ignoring the contamination of wgt.
By observing the results, the $p$-value of wgt is 0.008 for our proposed method,
which means wgt has significant influence on patients' survival time,
while the corresponding value is  0.244 for the naive method without the potential distorting effect of ``diam".
From the medical standpoint, wgt has great influence on patients' survival time,
whereas ignoring the contamination leads to this covariate insignificant.
Furthermore, from all these two methods, we can
conclude that patients with favorable tumor
would possess longer survival time, compared with ones with unfavorable tumor,
which coincides with Figure 1.
As a result, we conclude that the proposed covariate-calibration method
offers a convincing result for the Wilms' tumor data.

\section{Conclusion}
Covariate-adjusted problem  is a common contamination problem in biomedical studies.
Similar issues arrive in other field, e.g. in environmental studies,
exposures are often calibrated by the daily environment or
ambient measures, like the role of BMI in medical studies,
or genomic studies where library size is being normalized.
Our method deals with the type of some primary covariates that
are observed after being distorted by a multiplicative factor
(an unknown function of an observable confounding variable).
We fill in the gap in the literature on
censored survival data with distorting function in primary risk factor,
which is lacking in terms of statistical method.
We propose a direct estimation procedure to estimate the regression parameters
in the Cox proportional hazards regression model.
The novel idea of our procedure is to obtain a
consistent estimator of the distorted covariate
by employing the kernel smoothing method and then
obtain the  parameter estimation by plugging in the estimated covariate.

Numerical results show that the proposed
method is working very well in correcting the bias arising
from covariate  distortion.
It performs stably
to a variety choice of the distortion functions.
An important improvement of our method is that we allow
flexible distorting model to handle various confounding mechanisms.
The proposed method is easy to compute and will provide a critical tool
for researchers facing with this type data in practice.

A few remarks on using the proposed method in real studies.
First, on the construction of confidence interval of the proposed estimation,
we note that
because of the  nonlinear structure of the estimated partial likelihood
and the maximum  partial likelihood estimation does not have a closed form,
the establishment of theoretic properties in this paper is
more difficult than linear model.
The asymptotic covariance matrix derived in Theorem 1
depends on several unknown components,
therefore,
it is difficult to  construct confidence region based on normal approximation.
We recommend to use the common sandwich approach to obtain the standard error estimation.
This method has been tested and demonstrated to perform well in our numerical studies.

Second, for ease of exposition,
we consider only one  confounding variable.
In many applications, however,
there are multiple distorting variables that simultaneously affect
the primary covariate.
In principle,
the proposed method can handle this case
and the sandwich method can be employed as well to obtain the
standard error estimation.
Deriving theoretic properties of the corresponding estimators
will be more difficult and need  additional technicalities.

Finally, as we require division of the distorted variable by
the estimated distorting function,
we imposed some regularity assumptions on the curve of the distorting function.
In particular, the proposed method can not be applied if $E(X)$ vanishes.
Delaigle, Hall, \& Zhou (2016) proposed a more flexible
nonparametric estimator for the regression function,
which significantly weakens some of the
strong assumptions on the distorting function.
Further research is underway to extend this work to censored survival data.

\section{Supplementary materials}

The supplementary material presents the detailed proof of Theorem 1.

\section*{Acknowledgements}
This work is funded in part by  the National Natural Science Foundation of China NSFC 11771366 (Liu), NSFC 11971362 (Liu),
NSFC 11671311 (Wu) and NSFC 11901581 (Zhang),
U.S. National Institute of Health grant R01 ES021900 (Zhou), P01 CA142538 (Zhou), P30 ES010126 (Zhou).

\section*{Appendix A: Regularity conditions}
Unless otherwise stated, all limits are taken
as $n\rightarrow \infty$. Suppose ${\ba}=(a_1,\ldots,a_p)^{\rm T}$ and
${\bb}=(b_1,\ldots,b_p)^{\rm T}$ are $p$-vectors,
then we write ${\ba}\otimes {\bb}$ for the
matrix ${\bab}^{\rm T}$. Also we write ${\ba}^{\otimes 2}$
for the matrix ${\ba}\otimes {\ba}$. For a matrix $\bA$ or
vector $\ba$, let $\|{\bA}\|=\sup_{i,j}|a_{ij}|$ and
$\|{\ba}\|=\sup_{i}|a_i|$. For matrix or vector sequences $\bA_n$ and $\bB_n$, denote $\bA_n \stackrel{p}{\longrightarrow} \bA$ if $\| \bA_n - \bA\| \stackrel{p}{\longrightarrow} 0$ and denote $A_n = B_n + o_p(1)$ if $\| A_n - B_n \| \stackrel{p}{\longrightarrow} 0$. Denote $|{\ba}|=(\sum a_i^2)^{1/2}$
and diag$(\ba)$ as the diagonal matrix whose diagonal vector is $\ba$.
We set $\btheta = (\bbeta^{\rm T}, \gamma)^{\rm T}$, ${\bV}=({\bZ}^{\rm T},X)^{\rm T}$, $N_i(t)=I(\wt{T}_i\leq t, \Delta_i=1)$, $\bar{N}=\sum_{i=1}^nN_i$,
and $Y_i(t)=I(\wt{T}_i\geq t)$.
Let $\tau$ denote the end time of the study.
Here, we introduce the following notations:
\begin{eqnarray*}
S^{(l)}({\btheta},t)&=&\frac{1}{n}\displaystyle\sum_{i=1}^{n}{\bV}_i^{\otimes l}Y_i(t)\exp\left({\bV}_i^{\rm T}\btheta\right),\\
E({\btheta},t)&=&\frac{S^{(1)}({\btheta},t)}{S^{(0)}({\btheta},t)},\\
V({\btheta},t)&=&\frac{S^{(2)}({\btheta},t)}{S^{(0)}({\btheta},t)}-E({\btheta},t)^{\otimes 2},
\end{eqnarray*}
for $l=0,1,2$.
Note that $S^{(0)}({\btheta},t)$ is a scalar,
$S^{(1)}({\btheta},t)$ and $E({\btheta},t)$ are $(p+1)$-vectors,
$S^{(2)}({\btheta},t)$ and $V({\btheta},t)$ are $(p+1)\times (p+1)$ matrices.
Before proving the theorem, we first describe the
regular conditions needed as follows:
\begin{itemize}
\item[C1.] (Finite interval). $\int_0^\tau\lambda_0(t){\rm d}t < \infty$.
\item[C2.] (Asymptotic stability). There exist a neighbourhood
$\mathscr{B}$ of ${\btheta}_0$, scalar, vector and
            matrix functions $s^{(0)}$, $s^{(1)}$ and $s^{(2)}$
            defined on $\mathscr{B}\times [0,\tau]$ such that for $j=0,1,2,$
            \begin{eqnarray*}
            \sup_{t\in [0,\tau],{\btheta}\in \mathscr{B}}\|S^{(j)}({\btheta},t)-s^{(j)}({\btheta},t)\|\stackrel{p}{\longrightarrow} 0.
            \end{eqnarray*}
\item[C3.] (Lindeberg condition). There exists $\delta>0$ such that
            \begin{eqnarray*}
            n^{-1/2}\sup_{i,t}\left|{\bm V}_i\right|\,Y_i(t)\,I\left\{{\btheta}_0^{\rm T} {\bV}_i>-\delta\,|{\bV}_i|\right\}\stackrel{p}{\longrightarrow}0.
            \end{eqnarray*}
\item[C4.] (Asymptotic regularity conditions). Let $\mathscr{B}$,
           $s^{(0)}$, $s^{(1)}$ and $s^{(2)}$ be as in condition C$2$ and define
           $e=s^{(1)}/s^{(0)}$ and $v=s^{(2)}/s^{(0)}-e^{\otimes 2}$.
           For all ${\btheta}\in \mathscr{B}$, $t\in [0,\tau]$:
           $$s^{(1)}{({\btheta},t)}=\frac{\partial}{\partial {\btheta}}s^{(0)}{({\btheta},t)}, \;
           s^{(2)}{({\btheta},t)}=\frac{\partial^2}{\partial {\btheta}^2}s^{(0)}{({\btheta},t)},$$
           $s^{(0)}{(\cdot,t)}$, $s^{(1)}{(\cdot,t)}$ and $s^{(2)}{(\cdot,t)}$
           are continuous functions of ${\btheta}\in \mathscr{B}$,
           uniformly in $t\in [0,\tau]$, $s^{(0)}$, $s^{(1)}$ and $s^{(2)}$ are bounded on $\mathscr{B}\times [0,\tau]$, $s^{(0)}$ is bounded away from
           zero on $\mathscr{B}\times [0,\tau]$, and the matrix
            \begin{eqnarray*}
            \Sigma=\int_0^\tau v({\btheta}_0,t)s^{(0)}{({\btheta}_0,t)}\lambda_0(t)\,{\rm d}t
            \end{eqnarray*}
           is positive definite.
\item[C5.] $p(u)$ and $\phi(u)$ are bounded away from zero and have bounded second derivatives.
\item[C6.]
$\int_{-\infty}^\infty  K(x)\,{\rm d}x =  1$, $\int_{-\infty}^\infty x K(x)\,{\rm d}x =  0$ and $\int_{-\infty}^\infty x^2 K(x)\,{\rm d}x < \infty $.
\item[C7.]
The kernel function satisfies condition $K_1$ in Gin\'{e} \& Guillou (2002). Let
\bse
\mathscr{K} = \left\{ y \mapsto K(\frac{x - y}{h_n}) : x \in R, h_n > 0 \right\},
\ese
then for any $\epsilon > 0 $, $\mathscr{K}$ satisfies that
\bse
\sup_{P} N\left(\mathscr{K}, L_2(P), \epsilon \|F \|_{L_2(P)}\right) \le \left(  \frac{A}{\epsilon} \right)^\nu
\ese
for some positive constants $A$ and $\nu$, where $N(\Omega, d , \epsilon)$ denotes the $\epsilon$-covering number of the metric space $(\Omega , d)$, $F$ is the envelope function of $\mathscr{K}$, the supremum is taken over  $R$ and the norm $\| F\|^2_{L_2(P)}$ is defined as $\int_R |F(x)|^2{\rmd} P(x)$.
\item[C8.]
$|\log h_n|/\log\log n\rightarrow \infty$ and $nh_n/|\log h_n|\rightarrow \infty $;
$h_n$  and  $(nh_n)^{-1}$ monotonically converge to zero
as $n\rightarrow\infty$.
\item[C9.] $E(X)$ and $E({Z}_i)$ $(i=1,\ldots,p)$ are  bounded away from $0$.
\end{itemize}
These conditions are mild and can be satisfied in most of circumstances.
Conditions C1-C4 are essential for the asymptotic results
of Cox proportional hazards  regression model.
Condition C5 is a mild smoothness condition on the involved functions.
Condition C6 is  common for a kernel function and C7 is  to regularize the complexity of the kernel function so that the supremum
norm for kernel functions can be bounded in probability, which are also imposed in
Chen et al. (2016) and Chen, Genovese, \& Wasserman (2018). Specially, the Gauss kernel function satisfies the Conditions C6 and C7.
Condition C8  states that the bandwidth $h_n$
converges to zero at certain rate with respect to  the sample size $n$.
Condition C9 is necessary in the study of covariate-adjusted
problems, see Sent{\"u}rk \& M{\"u}ller (2006).

\section*{Appendix B: Proofs of asymptotic properties}
As a preparation, we state a lemma, which is extracted
from Lemma B.2 of Zhang, Zhu, \& Liang (2012) and frequently used in the process of the proof.
\begin{lemma}
Let $\eta(\bz)$ be a continuous function satisfying $E[\eta(\bZ)]^2<\infty$.
Assume that conditions C5$-$C9
hold. The following asymptotic representation holds:
\begin{eqnarray*}
\frac{1}{n}\ihe (\widehat X_i-X_i)\eta(\bZ_i) =
\frac{1}{n}\ihe(\wt X_i-X_i)\frac{E\l[X\eta(\bZ)\r]}{E(X)}+\textit{o}_p(n^{-1/2}).
\end{eqnarray*}
\end{lemma}

\subsection*{Proof of Theorem $1$}
{\it Proof of $(\romannumeral 1)$.} Denote by $\btheta = (\bbeta^{\rmT}, \gamma)^{\rm T},\;\bV=(\bZ^{\rmT},X)^{\rmT}$ and $\wh{\bV}=(\bZ^{\rmT},\wh X)^{\rmT}$,
the log partial likelihood of this covariate-adjusted Cox model can be written as
\begin{eqnarray*}
\wh L_n({\bbeta},\gamma)&=&\sum_{i=1}^{n} \int_{0}^{\tau} \wh{\bV}_i^{\rm T}\btheta\,\,{\rm d} N_i(t)
     - \int_{0}^{\tau} \log \left\{ \sum_{i=1}^{n} Y_i(t) \exp(\wh{\bV}_i^{\rm T}\btheta)\right\}\,\,{\rm d} \overline{N}(t).
\end{eqnarray*}
Set
\begin{eqnarray*}
 L_n({\bbeta},\gamma)&=&\sum_{i=1}^{n} \int_{0}^{\tau} {\bV}_i^{\rm T}\btheta\,\,{\rm d} N_i(t)
     - \int_{0}^{\tau} \log \left\{ \sum_{i=1}^{n} Y_i(t) \exp({\bV}_i^{\rm T}\btheta)\right\}\,\,{\rm d} \overline{N}(t).
\end{eqnarray*}
The main point of the proof lies in stating that, for any $\btheta\in \Theta$,
\begin{eqnarray*}
 \wh L_n({\bbeta},\gamma)- L_n({\bbeta},\gamma)=o_p(n).
\end{eqnarray*}
This implies, by the fact that
$\wh{\btheta}={\rm argmax}_{\btheta\in \Theta}\wh L_n({\bbeta},\gamma)$
and the consistency of Cox model under conditions C1--C4, the consistency of
$\wh{\btheta}$ follows from Lemma $1$ of Wu (1981).
The detailed proof were given in the supplementary material.

{\it Proof of $(\romannumeral 2)$.} Let
\begin{eqnarray*}
 \wh{U}(\btheta)=\sum_{i=1}^{n}\int_{0}^{\tau}\wh{\bV}_i\,\,{\rm d} N_i(t)
-\int_{0}^{\tau}\frac{\sum_{i=1}^{n} Y_i(t)\wh{\bV}_i\cdot\exp(\wh{\bV}_i^{\rm T}\btheta)}{\sum_{i=1}^{n} Y_i(t) \exp(\wh{\bV}_i^{\rm T}\btheta)}\,\,{\rm d} \overline{N}(t).
\end{eqnarray*}
By Taylor expansion, there exists  $\btheta^*$ between $\btheta_0$ and $\wh{\btheta}$
such that
\begin{eqnarray*}
\frac{1}{\sqrt{n}}  \wh{U}(\wh{\btheta})-\frac{1}{\sqrt{n}} \wh{U}(\btheta_0)&=&\frac{1}{n} \frac{\partial \wh{U}(\btheta^*) }{\partial \btheta}\sqrt{n}(\wh{\btheta}-\btheta_0).
\end{eqnarray*}
By the definition of $\wh{\btheta}$, we know that $\wh{U}(\wh{\btheta})=\bm 0$.
So we have
\begin{eqnarray*}
\sqrt{n}(\wh{\btheta}-\btheta_0)=\left\{-\frac{1}{n}\frac{\partial \wh{U}(\btheta^*) }{\partial \btheta}\right\}^{-1}\cdot\frac{1}{\sqrt{n}} \wh{U}(\btheta_0).
\end{eqnarray*}
We can prove that
\begin{eqnarray}\label{A4}
-\frac{1}{n}\frac{\partial \wh{U}({\btheta^*})}{\partial \btheta}\stackrel{p}{\longrightarrow}&\Sigma,
\end{eqnarray}
and
\begin{eqnarray}\label{A5}
\frac{1}{\sqrt n} \wh{ U}(\btheta_0)\stackrel{d}{\longrightarrow} \text{N}(0,\Sigma+\Omega),
\end{eqnarray}
where
$\Sigma=\begin{pmatrix}
  \Sigma_{11} & \Sigma_{12}\\
  \Sigma_{21} & \Sigma_{22}\\
 \end{pmatrix}$ is defined in condition C4,
 $ \zeta= (-\Sigma_{12}^{\rm T} \gamma_0, -\Sigma_{22} \gamma_0)^{\rm T}$ and
$\Omega=\frac{\text{Var}(\wt{X})-\text{Var}(X)}{\l\{E(X)\r\}^2} \zeta \zeta^{\rm T}$.

Combining (\ref{A4}) and (\ref{A5}), we have
\begin{eqnarray*}
\sqrt{n}(\wh{\btheta}-\btheta_0)
    &\stackrel{d}{\longrightarrow}&\text{N}(0,\Sigma^{-1}(\Sigma+\Omega)\Sigma^{-1}),
\end{eqnarray*}
where
\begin{eqnarray*}
\Sigma^{-1}(\Sigma+\Omega)\Sigma^{-1}&=&\Sigma^{-1}\Sigma\Sigma^{-1}+\frac{\text{Var}(\wt{X})-\text{Var}(X)}{\l\{E(X)\r\}^2}\Sigma^{-1}\zeta \zeta^{\rm T}\Sigma^{-1}\\
    &=&\Sigma^{-1}\Sigma\Sigma^{-1}+\frac{\text{Var}(\wt{X})-\text{Var}(X)}{\l\{E(X)\r\}^2}
         \left(\begin{array}{cccc}
         0&\cdots&0&0\\
         \vdots&&\vdots&0\\
         0&\cdots&0&0\\
         0&\cdots&0&\gamma_0^2\\
         \end{array}\right).
\end{eqnarray*}
We can obtain that
 \begin{eqnarray*}
           \sqrt n (\wh{\bbeta}-{\bbeta}_0)\stackrel{d}{\longrightarrow} \text{N}\l({\bf0},(\Sigma^{-1})_{p}\r),
  \end{eqnarray*}
and
 \begin{eqnarray*}
           \sqrt n (\wh{\gamma}-{\gamma}_0)\stackrel{d}{\longrightarrow} \text{N}\l({0},(\Sigma^{-1})_{(p+1,p+1)}+\frac{\text{Var}(\wt{X})-\text{Var}(X)}{\l\{E(X)\r\}^2}\gamma_0^2\r),
  \end{eqnarray*}
where $(\Sigma^{-1})_{p}$ and $(\Sigma^{-1})_{(p+1,p+1)}$
respectively represent the $p$th order sequential principal minor
and the $(p+1)$th diagonal element of matrix $\Sigma^{-1}$.
The detailed proof of (\ref{A4}) and (\ref{A5}) were given in the supplementary material.

%


\newpage
{\begin{table}[!h]
\small
\begin{center}
\caption{Simulation results for $\bbeta$ and $\gamma$  under sample size $n = 100$ }\label{tableex3}
\vspace{1ex}
{\setlength{\tabcolsep}{1.2mm}
\begin{tabular}{ccccccccccccccccccccc}

\hline

\hline

\hline
&&&\multicolumn{5}{c}{$\phi(u)={(u+3)}/{7}$}&&\multicolumn{5}{c}{$\phi(u)=3(u+1)^2/79$}\\
\cline{4-8}\cline{10-14}
CR&Method& Para. & Bias &SD&SE&MSE&CP&&Bias &SD&SE&MSE&CP\\
\hline
$20\%$&Proposed&$\beta_{1P}$& 0.020& 0.224& 0.222& 0.051& 0.960&&   0.006& 0.225& 0.221& 0.051& 0.946\\
      &        &$\beta_{2P}$& 0.007& 0.211& 0.206& 0.044& 0.942&&   0.001& 0.212& 0.206& 0.045& 0.943\\
      &        &$\gamma_{P}$& 0.021& 0.286& 0.278& 0.082& 0.946&&  -0.047& 0.292& 0.267& 0.087& 0.906\\\\
      &Naive   &$\beta_{1N}$&-0.006& 0.224& 0.221& 0.050& 0.945&&  -0.089& 0.223& 0.217& 0.058& 0.911\\
      &        &$\beta_{2N}$&-0.007& 0.211& 0.206& 0.044& 0.948&&  -0.045& 0.211& 0.205& 0.047& 0.933\\
      &        &$\gamma_{N}$&-0.217& 0.254& 0.235& 0.112& 0.780&&  -0.810& 0.171& 0.160& 0.686& 0.006\\\\
      &Oracle  &$\beta_{1O}$& 0.030& 0.222& 0.222& 0.050& 0.953&&   0.030& 0.222& 0.222& 0.050& 0.953\\
      &        &$\beta_{2O}$& 0.011& 0.210& 0.206& 0.044& 0.936&&   0.011& 0.210& 0.206& 0.044& 0.936\\
      &        &$\gamma_{O}$& 0.044& 0.276& 0.279& 0.078& 0.952&&   0.044& 0.276& 0.279& 0.078& 0.952\\
\\ \\
$40\%$&Proposed&$\beta_{1P}$& 0.024& 0.259& 0.253& 0.068& 0.944&&    0.011& 0.260& 0.252& 0.068& 0.941\\ &        &$\beta_{2P}$& 0.012& 0.251& 0.237& 0.063& 0.932&&    0.007& 0.251& 0.237& 0.063& 0.939\\
      &        &$\gamma_{P}$& 0.031& 0.334& 0.316& 0.113& 0.937&&   -0.042& 0.341& 0.303& 0.118& 0.897\\\\
      &Naive   &$\beta_{1N}$& 0.001& 0.258& 0.251& 0.067& 0.946&&   -0.073& 0.256& 0.247& 0.071& 0.920\\
      &        &$\beta_{2N}$& 0.000& 0.250& 0.237& 0.063& 0.942&&   -0.034& 0.250& 0.236& 0.063& 0.933\\
      &        &$\gamma_{N}$&-0.220& 0.299& 0.265& 0.137& 0.791&&   -0.824& 0.200& 0.180& 0.718& 0.026\\\\
      &Oracle  &$\beta_{1O}$& 0.034& 0.257& 0.253& 0.067& 0.942&&    0.034& 0.257& 0.253& 0.067& 0.942\\
      &        &$\beta_{2O}$& 0.015& 0.250& 0.237& 0.063& 0.928&&    0.015& 0.250& 0.237& 0.063& 0.928\\
      &        &$\gamma_{O}$& 0.050& 0.324& 0.315& 0.107& 0.951&&    0.050& 0.324& 0.315& 0.107& 0.951\\
\\ \\
$80\%$&Proposed&$\beta_{1P}$& 0.072& 0.454 &0.436& 0.212& 0.932&&   0.065& 0.461& 0.435& 0.217& 0.931\\
      &        &$\beta_{2P}$& 0.031& 0.454 &0.413& 0.207& 0.914&&   0.028& 0.452& 0.412& 0.205& 0.912\\
      &        &$\gamma_{P}$& 0.107& 0.589 &0.542& 0.359& 0.939&&   0.020& 0.572& 0.517& 0.328& 0.927\\\\
      &Naive   &$\beta_{1N}$& 0.059& 0.466 &0.434& 0.221& 0.930&&   0.012& 0.460& 0.428& 0.212& 0.927\\
      &        &$\beta_{2N}$& 0.022& 0.450 &0.413& 0.203& 0.913&&   0.002& 0.449& 0.410& 0.201& 0.915\\
      &        &$\gamma_{N}$&-0.183& 0.507 &0.449& 0.291& 0.883&&  -0.830& 0.340& 0.304& 0.805& 0.254\\\\
      &Oracle  &$\beta_{1O}$& 0.080& 0.460 &0.437& 0.218& 0.935&&   0.080& 0.460& 0.437& 0.218& 0.935\\
      &        &$\beta_{2O}$& 0.033& 0.458 &0.414& 0.211& 0.909&&   0.033& 0.458& 0.414& 0.211& 0.909\\
      &        &$\gamma_{O}$& 0.125& 0.592 &0.539& 0.366& 0.939&&   0.125& 0.592& 0.539& 0.366& 0.939\\
\hline
\end{tabular}
}
\end{center}
\vspace{0ex}
{The true value of the parameters
$\beta_{1}=1$, $\beta_{2}=0.5$, $\gamma=1.5$;
$\phi(\cdot)$, the distortion function;
CR, the censoring rate;
Bias, the estimate value minus the true value;
SD, the standard deviation;
SE, the estimate of SD;
MSE, the mean-square error;
CP, empirical coverage percentage of the $95\%$ confidence interval.
}
\end{table}
}

\newpage
{\begin{table}[!h]
\small
\begin{center}
\caption{Simulation results for $\bbeta$ and $\gamma$  under sample size $n = 200$ }\label{tableex3}
\vspace{1ex}
{\setlength{\tabcolsep}{1.2mm}
\begin{tabular}{ccccccccccccccccccccc}

\hline

\hline

\hline
&&&\multicolumn{5}{c}{$\phi(u)={(u+3)}/{7}$}&&\multicolumn{5}{c}{$\phi(u)=3(u+1)^2/79$}\\
\cline{4-8}\cline{10-14}
CR&Method& Para. & Bias &SD&SE&MSE&CP&&Bias &SD&SE&MSE&CP\\
\hline
$20\%$&Proposed&$\beta_{1P}$& 0.009& 0.156& 0.151& 0.024& 0.939&&   0.004& 0.157& 0.151& 0.025& 0.936\\
      &        &$\beta_{2P}$& 0.004& 0.151& 0.142& 0.023& 0.936&&   0.002& 0.152& 0.142& 0.023& 0.935\\
      &        &$\gamma_{P}$& 0.006& 0.196& 0.190& 0.039& 0.949&&  -0.021& 0.202& 0.186& 0.041& 0.932\\
      \\
      &Naive   &$\beta_{1N}$&-0.017& 0.158& 0.150& 0.025& 0.932&&  -0.094& 0.158& 0.147& 0.034& 0.865\\
      &        &$\beta_{2N}$&-0.009& 0.152& 0.141& 0.023& 0.930&&  -0.049& 0.153& 0.140& 0.026& 0.907\\
      &        &$\gamma_{N}$&-0.234& 0.178& 0.159& 0.087& 0.649&&  -0.812& 0.122& 0.108& 0.674& 0.000\\
      \\
      &Oracle  &$\beta_{1O}$& 0.015& 0.154& 0.151& 0.024& 0.935&&   0.015& 0.154& 0.151& 0.024& 0.935\\
      &        &$\beta_{2O}$& 0.007& 0.150& 0.142& 0.022& 0.939&&   0.007& 0.150& 0.142& 0.022& 0.939\\
      &        &$\gamma_{O}$& 0.021& 0.190& 0.190& 0.036& 0.948&&   0.021& 0.190& 0.190& 0.036& 0.948\\
\\ \\
$40\%$&Proposed&$\beta_{1P}$& 0.015& 0.177& 0.172& 0.032& 0.941&&  0.010 &0.178& 0.172 &0.032& 0.938\\
      &        &$\beta_{2P}$& 0.002& 0.169& 0.163& 0.029& 0.946&& -0.001 &0.170& 0.163 &0.029& 0.945\\
      &        &$\gamma_{P}$& 0.015& 0.225& 0.215& 0.051& 0.944&& -0.013 &0.228& 0.211 &0.052& 0.928\\
      \\
      &Naive   &$\beta_{1N}$&-0.007& 0.180& 0.171& 0.032& 0.929&& -0.074 &0.181& 0.168 &0.038& 0.892\\
      &        &$\beta_{2N}$&-0.010& 0.171& 0.162& 0.029& 0.936&& -0.046 &0.172& 0.161 &0.032& 0.921\\
      &        &$\gamma_{N}$&-0.235& 0.199& 0.179& 0.095& 0.698&& -0.822 &0.134& 0.122 &0.693& 0.001\\
      \\
      &Oracle  &$\beta_{1O}$& 0.021& 0.175& 0.172& 0.031& 0.940&&  0.021 &0.175& 0.172 &0.031& 0.940\\
      &        &$\beta_{2O}$& 0.005& 0.168& 0.163& 0.028& 0.945&&  0.005 &0.168& 0.163 &0.028& 0.945\\
      &        &$\gamma_{O}$& 0.028& 0.220& 0.215& 0.049& 0.948&&  0.028 &0.220& 0.215 &0.049& 0.948\\
\\ \\
$80\%$&Proposed&$\beta_{1P}$& 0.022& 0.294& 0.293& 0.087& 0.957&&  0.020& 0.294& 0.293& 0.087& 0.958\\
      &        &$\beta_{2P}$& 0.014& 0.285& 0.283& 0.081& 0.952&&  0.012& 0.285& 0.282& 0.082& 0.948\\
      &        &$\gamma_{P}$& 0.023& 0.383& 0.360& 0.147& 0.934&& -0.010& 0.383& 0.353& 0.147& 0.914\\
      \\
      &Naive   &$\beta_{1N}$& 0.011& 0.294& 0.292& 0.087& 0.951&& -0.028& 0.293& 0.289& 0.087& 0.943\\
      &        &$\beta_{2N}$& 0.006& 0.285& 0.282& 0.081& 0.943&& -0.018& 0.285& 0.280& 0.082& 0.941\\
      &        &$\gamma_{N}$&-0.247& 0.327& 0.298& 0.168& 0.816&& -0.850& 0.223& 0.202& 0.773& 0.035\\
      \\
      &Oracle  &$\beta_{1O}$& 0.026& 0.293& 0.293& 0.086& 0.955&&  0.026& 0.293& 0.293& 0.086& 0.955\\
      &        &$\beta_{2O}$& 0.017& 0.284& 0.282& 0.081& 0.951&&  0.017& 0.284& 0.282& 0.081& 0.951\\
      &        &$\gamma_{O}$& 0.036& 0.376& 0.360& 0.143& 0.941&&  0.036& 0.376& 0.360& 0.143& 0.941\\
\hline
\end{tabular}
}
\end{center}
\vspace{0ex}
{The true value of the parameters
$\beta_{1}=1$, $\beta_{2}=0.5$, $\gamma=1.5$;
$\phi(\cdot)$, the distortion function;
CR, the censoring rate;
Bias, the estimate value minus the true value;
SD, the standard deviation;
SE, the estimate of SD;
MSE, the mean-square error;
CP, empirical coverage percentage of the $95\%$ confidence interval.
}
\end{table}
}

{\begin{table}[!h]
\caption{The data of the NWTSG trials grouped by
the histological type }\label{table3}
\begin{center}
\vspace{0.5ex}
{\setlength{\tabcolsep}{3mm}
\begin{tabular}{lrrr}
\hline

\hline

\hline

    &overall& favorable& unfavorable  \\
\hline
size& $3915$  & $3476$ &$439$ \\
wgt&$604.56 $& $603.74 $& $611.12 $\\
diam& $11.21$& $11.20$ & $11.32$\\
age& $3.53$& $3.52$& $3.68 $\\
stage$(\%)$& $64.78$ & $66.28$ &$52.85$\\
num$(\%)$& $42.68$ & $42.55$ & $43.74$\\
time & $10.33 $ & $10.68 $ & $7.55$\\
cen.rate $(\%)$& $88.66$ & $92.23$ & $60.36$\\

\hline

\hline

\hline
\end{tabular}
}
\end{center}
{
overall, the total patients;
favorable, the patients with favorable tumor;
unfavorable, the patients with unfavorable tumor;
size, the sample size;
wgt, the mean weight of tumor bearing specimens;
diam, the mean diameter of tumors;
age, the mean age of patients at diagnosis;
stage, the percentage of patients with tumor localized spread;
num,  the percentage of patients in NWTS-3 trial;
time, the mean observed  time;
cen.rate, the censoring rate.
}
\end{table}
}

{
\begin{table}[!h]
\caption{The analysis results of the covariate effects in the NWTSG trials}\label{table4}
\begin{center}
\vspace{0.5ex}
{\setlength{\tabcolsep}{5mm}
\begin{tabular}{ccrrrr}
\hline

\hline

\hline
Method&Covariate&EST& SE&$P$-value  \\
\hline

     Proposed&wgt  &$              -0.482$& $               0.180$&   $0.008$ \\
             &type &$               1.820$& $               0.096$&   $<0.001$ \\
             &stage&$              -0.900$& $               0.097$&   $<0.001$ \\
             &age  &$               0.070$& $               0.020$&   $<0.001$ \\
             &num  &$               0.171$& $               0.098$&   $0.081$ \\
  \\
       Naive  &wgt  &$             -0.139$& $               0.119$&   $0.244$  \\
             &type &$               1.821$& $               0.096$&   $<0.001$ \\
             &stage&$              -0.908$& $               0.099$&   $<0.001$ \\
             &age  &$               0.066$& $               0.020$&   $0.001$  \\
             &num  &$               0.187$& $               0.097$&   $0.055$ \\
\hline

\hline

\hline
\end{tabular}
}
\end{center}
{wgt, the weight of tumor bearing specimen;
type, the histological type of the tumor;
stage, the tumor stage;
age, the age of patients at diagnosis;
num, the study number;
EST, the estimate of the parameters;
SE, the standard error estimate;
$P$-value, the $p$-value of the parameters.
}
\end{table}
}

\begin{figure}[!h]
\centering\includegraphics[width=14cm]{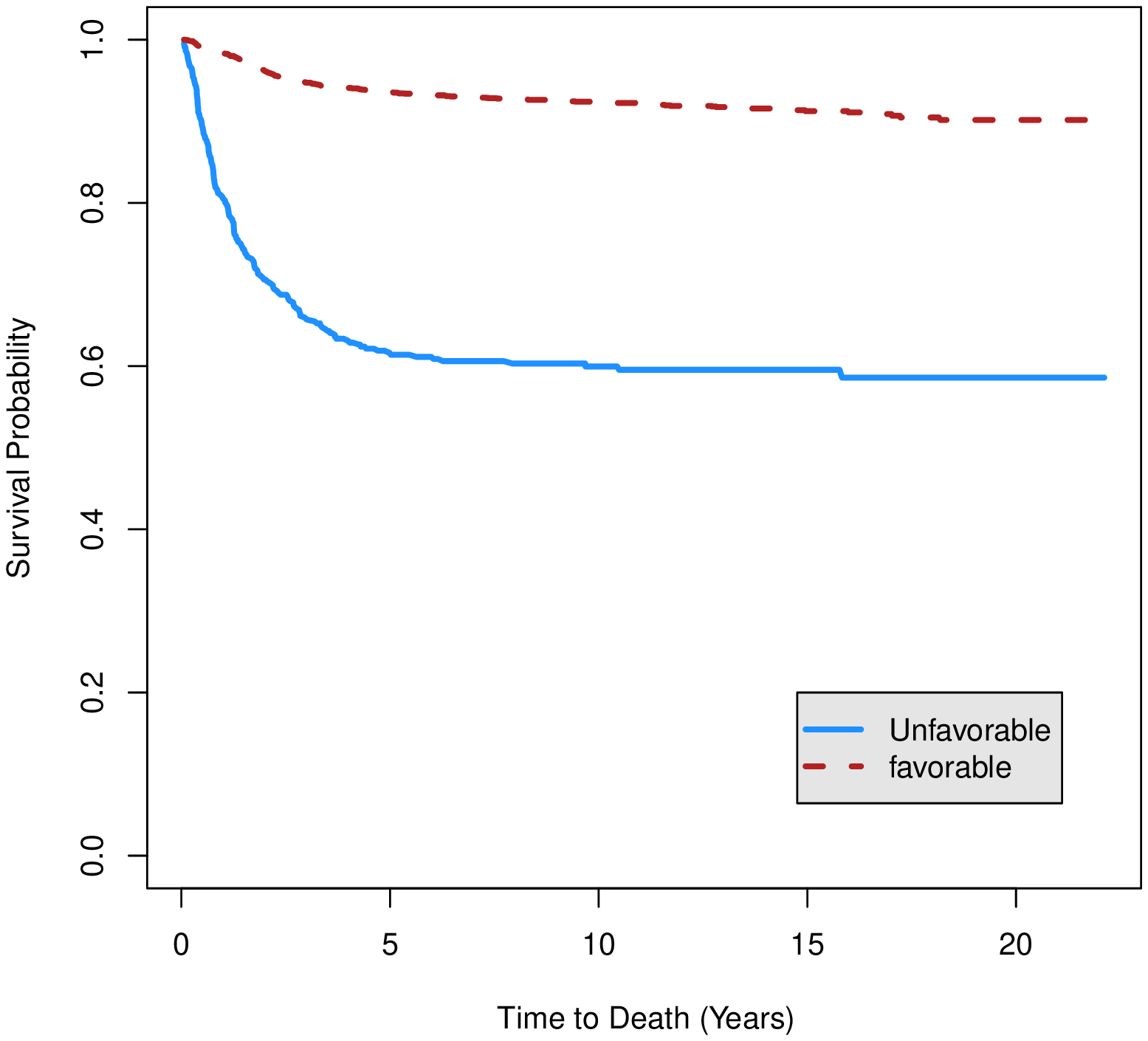}
  \caption{Kaplan-Meier survival curves stratified by
  two different histological types of the tumor in in the NWTSG trials.}
 \label{fig-KM}
\end{figure}

\begin{figure}[!h]
\centering\includegraphics[width=14cm]{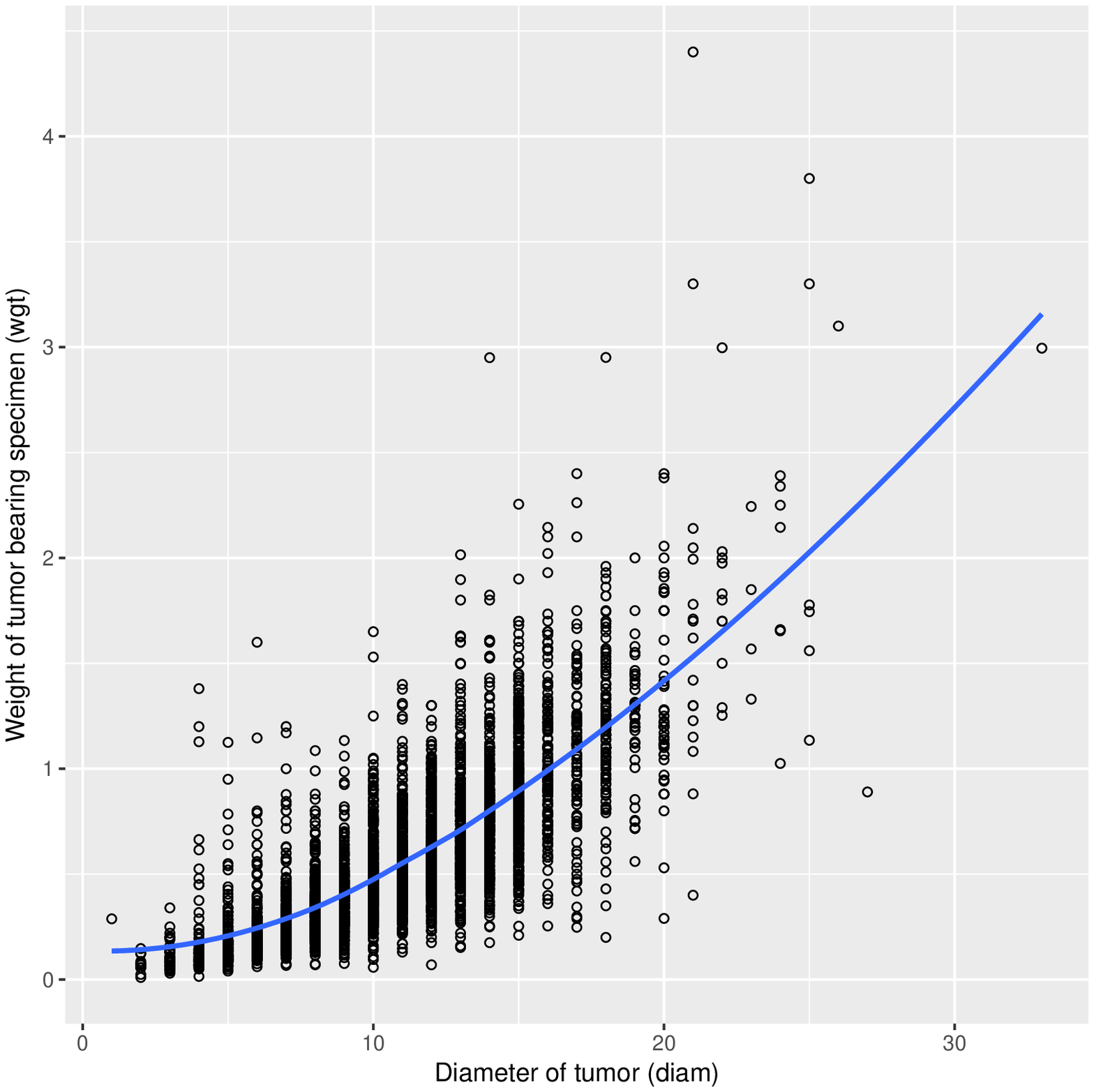}
  \caption{The scatter diagram of tumor bearing specimen's weight (wgt)
  versus tumor's diameter (diam) for the NWTSG trials.}
 \label{fig-KM}
\end{figure}

\end{document}


\begin{frontmatter}

\title{Supplementary material for ``Cox regression analysis for distorted covariates with an unknown distortion function"}

\author[a]{Yanyan Liu}
\author[b]{Yuanshan Wu}
\author[b]{Jing Zhang\corref{cor}}
\author[c]{Haibo Zhou\corref{cor}}

\address[a]{School of Mathematics and Statistics, Wuhan University, Wuhan, Hubei 430072, China}
\address[b]{School of Statistics and Mathematics, Zhongnan University of Economics and Law, Wuhan, Hubei 430073, China}
\address[c]{Department of Biostatistics, University of North Carolina at Chapel Hill, Chapel Hill, NC
27599-7420, USA}

\cortext[cor]
{Corresponding author: jing66@zuel.edu.cn(Jing Zhang), zhou@bios.unc.edu(Haibo Zhou)}

\end{frontmatter}

This supplementary material contains the detailed proof of Theorem 1.

\section*{Proof of Theorem $1$}

As a preparation, we state a lemma, which is extracted
from Lemma B.2 of Zhang, Zhu, \& Liang (2012) and frequently used in the process of the proof.
\begin{lemma}
Let $\eta(\bz)$ be a continuous function satisfying $E[\eta(\bZ)]^2<\infty$.
Assume that conditions C5$-$C9
hold. The following asymptotic representation holds:
\begin{eqnarray*}
\frac{1}{n}\ihe (\wh{X}_i-X_i)\eta(\bZ_i) =
\frac{1}{n}\ihe(\wt{X}_i-X_i)\frac{E\l[X\eta(\bZ)\r]}{E(X)}+\textit{o}_p(n^{-1/2}).
\end{eqnarray*}
\end{lemma}

{\it Proof of $(\romannumeral 1)$.} Denote by $\btheta = (\bbeta^{\rmT}, \gamma)^{\rm T},\;\bV=(\bZ^{\rmT},X)^{\rmT}$ and $\wh{\bV}=(\bZ^{\rmT},\wh X)^{\rmT}$,
the log partial likelihood of this covariate-adjusted Cox model can be written as
\begin{eqnarray*}
\wh L_n({\bbeta},\gamma)&=&\sum_{i=1}^{n} \int_{0}^{\tau} \wh{\bV}_i^{\rm T}\btheta\,\,{\rm d} N_i(t)
     - \int_{0}^{\tau} \log \left\{ \sum_{i=1}^{n} Y_i(t) \exp(\wh{\bV}_i^{\rm T}\btheta)\right\}\,\,{\rm d} \overline{N}(t).
\end{eqnarray*}
Set
\begin{eqnarray*}
 L_n({\bbeta},\gamma)&=&\sum_{i=1}^{n} \int_{0}^{\tau} {\bV}_i^{\rm T}\btheta\,\,{\rm d} N_i(t)
     - \int_{0}^{\tau} \log \left\{ \sum_{i=1}^{n} Y_i(t) \exp({\bV}_i^{\rm T}\btheta)\right\}\,\,{\rm d} \overline{N}(t),
\end{eqnarray*}
The main point of the proof lies in stating that, for any $\btheta\in \Theta$,
\begin{eqnarray*}
 \wh L_n({\bbeta},\gamma)- L_n({\bbeta},\gamma)=o_p(n).
\end{eqnarray*}
This implies, by the fact that
$\wh{\btheta}={\rm argmax}_{\btheta\in \Theta}\wh L_n({\bbeta},\gamma)$
and the consistency of Cox model under conditions C1--C4, the consistency of
$\wh{\btheta}$ follows from Lemma $1$ of Wu (1981).
Now, after simple calculations, we can obtain the decomposition
\begin{eqnarray}\label{A1}
&         &  \wh L_n({\bbeta},\gamma)- L_n({\bbeta},\gamma)\nonumber\\
&=         & \sum_{i=1}^{n} \int_{0}^{\tau} (\gamma \widehat{X}_i - \gamma X_i) \,{\rm d} N_i(t)
            + \int_{0}^{\tau} \log \left\{\frac{\sum_{i=1}^{n} Y_i(t)\exp({\bV}_i^{\rm T}\bm\theta)}{\sum_{i=1}^{n} Y_i(t)
             \exp(\wh{\bV}_i^{\rm T}\btheta)}\right\}\,{\rm d} \overline{N}(t)\nonumber\\
&\triangleq& F_1 + F_2,
\end{eqnarray}
where
\begin{eqnarray*}
F_1 =\sum_{i=1}^{n} \int_{0}^{\tau} (\gamma \widehat{X}_i - \gamma X_i) \,{\rm d} N_i(t),
\end{eqnarray*}
and
\begin{eqnarray*}
F_2 =\int_{0}^{\tau} \log \left\{\frac{\sum_{i=1}^{n} Y_i(t)\exp({\bV}_i^{\rm T}\bm\theta)}{\sum_{i=1}^{n} Y_i(t)
             \exp(\wh{\bV}_i^{\rm T}\btheta)}\right\}\,{\rm d} \overline{N}(t).
\end{eqnarray*}
Define
\begin{eqnarray*}
g_{X}(U)=E(\wt{X}|U)p(U),
\end{eqnarray*}
then
\begin{eqnarray*}
\wh{g}_{X}(U)=\frac{1}{nh_n}\sum_{i=1}^n K\left(\frac{u-U_i}{h_n}\right)\wt{X}_i.
\end{eqnarray*}
It follows from Theorem 2 in Chen,  Genovese, \& Wasserman (2014) that
\bsee\label{mm1}
\sup_u |\hat g_X(u) - g_X(u)| = O_P\left(h_n^2+\sqrt{\frac{\log n}{nh_n}} \right).
\esee
By Theorem 2.3 in Gin\'{e} \& Guillou (2002), we have
\bse
\sup_{u}\left|\hat{p}(u) - E\{\hat{p}(u)\}\right| = O\left(\sqrt{\frac{\log n}{nh_n}} \right), \quad a.s..
\ese
On the other hand, under assumption C6, it is well known that
\bse
\sup_{u}\left| E\{\hat{p}(u)\} - p(u) \right| = O(h_n^2).
\ese
As a consequence, we have
\bsee\label{mm2}
\sup_{u}\left|\hat{p}(u) - p(u)\right| = O\left(h_n^2+\sqrt{\frac{\log n}{nh_n}} \right), \quad a.s..
\esee
Coupled with (\ref{mm1}) and (\ref{mm2}), some straightforward calculations entail
\begin{eqnarray}\label{A2}
|F_1| &=& \left|\sum_{i=1}^{n} \int_{0}^{\tau} \gamma (\widehat{X}_i - X_i) \,{\rm d} N_i(t)\right|  \nonumber\\
    &=& O_p\left\{ \left(h_n^2 + (nh_n)^{-1/2}(\log n)^{1/2} \right)\cdot n \right\} \nonumber\\
    &=& o_p(n).
\end{eqnarray}
Similarly,
\begin{eqnarray}\label{A3}
|F_2| &=& 
 \left|\int_{0}^{\tau} \log \left\{\frac{\frac{1}{n}\sum_{i=1}^{n} Y_i(t) \exp(\bbeta^{\rm T} {\bZ}_i)\exp(\gamma X_i)}
         {\frac{1}{n}\sum_{i=1}^{n} Y_i(t) \exp(\bbeta^{\rm T} {\bZ}_i)\exp(\gamma \widehat{X}_i)}\right\}\,{\rm d} \overline{N}(t)\right|\nonumber
 \\
&=& O_p\left\{ \left(h_n^2 + (nh_n)^{-1/2}(\log n)^{1/2} \right)\cdot n \right\} \nonumber\\
    &=& o_p(n).
\end{eqnarray}
Combining (\ref{A1}), (\ref{A2}) and (\ref{A3}), we have
\begin{eqnarray*}
 \wh L_n({\bbeta},\gamma)- L_n({\bbeta},\gamma)=o_p(n).
\end{eqnarray*}
Here, we complete the proof of $(\romannumeral 1)$.
\\
{\it Proof of $(\romannumeral 2)$.} Let
\begin{eqnarray*}
 \wh{U}(\btheta)=\sum_{i=1}^{n}\int_{0}^{\tau}\wh{\bV}_i\,\,{\rm d} N_i(t)
-\int_{0}^{\tau}\frac{\sum_{i=1}^{n} Y_i(t)\wh{\bV}_i\cdot\exp(\wh{\bV}_i^{\rm T}\btheta)}{\sum_{i=1}^{n} Y_i(t) \exp(\wh{\bV}_i^{\rm T}\btheta)}\,\,{\rm d} \overline{N}(t).
\end{eqnarray*}
By Taylor expansion, there exists  $\btheta^*$ between $\btheta_0$ and $\wh{\btheta}$
such that
\begin{eqnarray*}
\frac{1}{\sqrt{n}}  \wh{U}(\wh{\btheta})-\frac{1}{\sqrt{n}} \wh{U}(\btheta_0)&=&\frac{1}{n} \frac{\partial \wh{U}(\btheta^*) }{\partial \btheta}\sqrt{n}(\wh{\btheta}-\btheta_0).
\end{eqnarray*}
By the definition of $\wh{\btheta}$, we know that $\wh{U}(\wh{\btheta})=\bm 0$.
So we have
\begin{eqnarray*}
\sqrt{n}(\wh{\btheta}-\btheta_0)=\left\{-\frac{1}{n}\frac{\partial \wh{U}(\btheta^*) }{\partial \btheta}\right\}^{-1}\cdot\frac{1}{\sqrt{n}} \wh{U}(\btheta_0).
\end{eqnarray*}
In order to prove the asymptotic normality, it suffices to  show that
\begin{eqnarray*}
\left\{-\frac{1}{n}\frac{\partial \wh{U}(\btheta^*) }{\partial \btheta}\right\}^{-1}\stackrel{p}{\longrightarrow} \text{a\  non-singular\  matrix},
\end{eqnarray*}
and
\begin{eqnarray*}
\frac{1}{\sqrt{n}} \wh{U}(\btheta_0)\stackrel{d}{\longrightarrow} \text{a\  Gaussian\  distribution}.
\end{eqnarray*}
We introduce some notations:
\begin{eqnarray*}
\widehat{S}^{(0)}({\btheta},t)&=&\frac{1}{n}\displaystyle\sum_{i=1}^{n}Y_i(t)\exp\left(\wh{\bV}_i^{\rm T}\btheta\right),\\
S_{X}^{(l)}({\btheta},t)&=&\frac{1}{n}\sum_{i=1}^{n} Y_i(t) X_i^{\otimes l}\exp\left({\bV}_i^{\rm T}\btheta\right),\\
\widehat{S}_{X}^{(l)}({\btheta},t)&=& \frac{1}{n}\sum_{i=1}^{n} Y_i(t) \wh{X}_i^{\otimes l}\exp\left(\wh{\bV}_i^{\rm T}\btheta\right),\\
S_{\bZ}^{(l)}({\btheta},t)&=&\frac{1}{n}\sum_{i=1}^{n} Y_i(t) {\bZ}_i^{\otimes l}\exp\left({\bV}_i^{\rm T}\btheta\right),\\
\widehat{S}_{\bZ}^{(l)}({\btheta},t)&=&\frac{1}{n}\sum_{i=1}^{n} Y_i(t) {\bZ}_i^{\otimes l}\exp\left(\wh{\bV}_i^{\rm T}\btheta\right),\\
S_{\bZ X}^{(l_1,l_2)}({\btheta},t)&=&\frac{1}{n}\sum_{i=1}^{n} Y_i(t) {\bZ}_i^{\otimes l_1} X_i^{\otimes l_2} \exp\left({\bV}_i^{\rm T}\btheta\right),\\
\widehat{S}_{\bZ X}^{(l_1,l_2)}({\btheta},t)&=&\frac{1}{n}\sum_{i=1}^{n} Y_i(t) {\bZ}_i^{\otimes l_1} \wh{X}_i^{\otimes l_2} \exp\left(\wh{\bV}_i^{\rm T}\btheta\right),
\end{eqnarray*}
$s_{X}^{(l)}({\btheta},t)$, $s_{\bZ}^{(l)}({\btheta},t)$ and
$s_{\bZ X}^{(l_1,l_2)}({\btheta},t)$ are defined the same as $s^{(0)}$, $s^{(1)}$
and $s^{(2)}$ in condition C$2$, where $l,l_1,l_2=1,2,3$.

Straightforward calculations entail
\begin{eqnarray*}
    & &\left\|\wh{S}^{(0)}({\btheta},t) - {S}^{(0)}({\btheta},t)\right\| \\
    &=& \left\|\frac{1}{n} \sum_{i=1}^{n} Y_i(t) \exp(\bbeta^{\rm T} \bZ_i)\left\{\exp(\gamma \widehat{X}_i)- \exp(\gamma X_i)\right\}\right\| \\
    &=& \left\|\frac{1}{n} \sum_{i=1}^{n} Y_i(t) \exp(\bbeta^{\rm T} \bZ_i)
           \left\{\gamma\exp(\gamma X_i)(\wh{X}_i-X_i)+\frac{1}{2}\gamma^2\exp(\gamma X_i^{*})(\wh{X}_i-X_i)^2\right\}\right\|\\
    &\le &\left\|\frac{1}{n} \sum_{i=1}^{n} Y_i(t) \exp\left( {\bV}_i^{\rm T}\btheta \right)\cdot\gamma\cdot(\wh{X}_i-X_i)\right\|+o_p\left(n^{-1/2}\right) \\
    &\le& \left\| \frac{1}{n}\sum_{i=1}^{n}(\wt{X}_i-X_i) \frac{E\left\{XY(t)\exp({\bV}^{\rm T}\btheta)\gamma\right\}}{E(X)}\right\|+o_p(n^{-1/2})+ o_p\left(n^{-1/2}\right)\\
    &=& \left\|\sum_{i=1}^{n}(\wt{X}_i-X_i) \frac{s_{X}^{(1)}(\btheta,t)\cdot\gamma}{nE(X)} \right\|+ o_p\left(n^{-1/2}\right),
\end{eqnarray*}
where $X_i^* = \hat{X}_i + t_i^*(\hat{X}_i - X_i)$ for some $t^*_i \in ( 0 , 1)$.
Set $Q_n = \frac{\sum_{i=1}^{n}(\wt{X}_i-X_i) }{E(X)}$,
then
\bse
\sup_{\btheta \in \mathscr{B} , t \in [ 0 , \tau ]}\left|\widehat{S}^{(0)}({\btheta},t) - {S}^{(0)}({\btheta},t) \right| \le \sup_{\btheta \in \mathscr{B} , t \in [ 0 , \tau ]}\left|\frac{s_{X}^{(1)}(\btheta,t)\cdot\gamma}{n}Q_n\right|
+o_p\left(n^{-1/2}\right)= o_p(1).
\ese
Likewise, it holds uniformly over $(\btheta, t) \in \mathscr{B} \times [0 , \tau]$ that
\begin{eqnarray*}
\left\|\widehat{S}_{\bZ}^{(1)}({\btheta},t) - S_{\bZ}^{(1)}({\btheta},t)\right\|
& \le &
\left\|\frac{s_{\bZ X}^{(1,1)}({\btheta},t)\cdot\gamma}{n}Q_n \right\|+ o_p\left(n^{-1/2}\right)
=o_p(1),
\\
\left\| \widehat{S}_{{X}}^{(1)}(\btheta,t) - S_{X}^{(1)}(\btheta,t)\right\|
& \le &
\left\| \frac{s_{X}^{(1)}(\btheta,t)}{n}Q_n \right\| +
\left\| \frac{\gamma \cdot s_{X}^{(2)}(\btheta,t)}{n}Q_n \right\|+ o_p\left(n^{-1/2}\right)
=o_p(1),
\\
\left\| \widehat{S}_{{\bZ}}^{(2)}(\btheta,t) - S_{\bm Z}^{(2)}(\btheta,t)\right\|
& \le &
\left\| \frac{\gamma \cdot s_{\bZ X}^{(2,1)}(\btheta,t)}{n}Q_n \right\|+ o_p\left(n^{-1/2}\right)
=o_p(1),
\\
\left\| \widehat S_{{X}}^{(2)}(\btheta,t) - S_{X}^{(2)}(\btheta,t)\right\|
&\le &
\left\| \frac{2s_{X}^{(2)}(\btheta,t)}{n}Q_n\right\| +
\left\| \frac{\gamma s_{X}^{(3)}(\btheta,t)}{n}Q_n \right\|+ o_p\left(n^{-1/2}\right)
=o_p(1),
\\
\left\| \widehat {S}_{\bZ{X}}^{(1,1)}(\btheta,t) - S_{{\bZ}X}^{(1,1)}(\btheta,t)\right\|
& \le &
\left\| \frac{s_{\bZ X}^{(1,1)}(\btheta,t)}{n}Q_n \right\| +
\left\| \frac{\gamma s_{\bZ X}^{(1,2)}(\btheta,t)}{n}Q_n \right\|+ o_p\left(n^{-1/2}\right)
=o_p(1).
\end{eqnarray*}

For simplicity, we define
\begin{eqnarray*}
\widehat{G}(\btheta,t)=
\left( \begin{array}{cc}
\frac{\widehat{S}_{\bZ}^{(2)}(\btheta,t)}{\widehat{S}^{(0)}(\btheta,t)}
-\frac{\widehat{S}_{\bZ}^{(1)}(\btheta,t)\cdot \left\{\widehat{S}_{\bZ}^{(1)}(\btheta,t)\right\}^{\rm T} }{\left\{\widehat{S}^{(0)}(\btheta,t)\right\}^2},
& \frac{\widehat{S}_{\bZ{X}}^{(1,1)}(\btheta,t)}{\widehat{S}^{(0)}(\btheta,t)}
-\frac{\widehat{S}_{\bZ}^{(1)}(\btheta,t)\cdot \widehat{S}_{{X}}^{(1)}(\btheta,t)}{\left\{\widehat{S}^{(0)}(\btheta,t)\right\}^2}\\
\left[
 \frac{\widehat{S}_{\bZ{X}}^{(1,1)}(\btheta,t)}{\widehat{S}^{(0)}(\btheta,t)}
-\frac{\widehat{S}_{\bZ}^{(1)}(\btheta,t)\cdot \widehat{S}_{{X}}^{(1)}(\btheta,t)}{\left\{\widehat{S}^{(0)}(\btheta,t)\right\}^2}
\right]^{\rm T},
& \frac{\widehat{S}_{{X}}^{(2)}(\btheta,t)}{\widehat{S}^{(0)}(\btheta,t)}- \frac{\left\{\widehat{S}_{{X}}^{(1)}(\btheta,t)\right\}^2}{\left\{\widehat{S}^{(0)}(\btheta,t)\right\}^2}\\
\end{array}
 \right)
\end{eqnarray*}
and
\begin{eqnarray*}
{G}(\btheta,t)=
\left( \begin{array}{cc}
\frac{{S}_{\bZ}^{(2)}(\btheta,t)}{{S}^{(0)}(\btheta,t)}
-\frac{{S}_{\bZ}^{(1)}(\btheta,t)\cdot \left\{{S}_{\bZ}^{(1)}(\btheta,t)\right\}^{\rm T} }{\left\{{S}^{(0)}(\btheta,t)\right\}^2},
& \frac{{S}_{\bZ{X}}^{(1,1)}(\btheta,t)}{{S}^{(0)}(\btheta,t)}
-\frac{{S}_{\bZ}^{(1)}(\btheta,t)\cdot {S}_{{X}}^{(1)}(\btheta,t)}{\left\{{S}^{(0)}(\btheta,t)\right\}^2}\\
\left[
 \frac{{S}_{\bZ{X}}^{(1,1)}(\btheta,t)}{{S}^{(0)}(\btheta,t)}
-\frac{{S}_{\bZ}^{(1)}(\btheta,t)\cdot {S}_{{X}}^{(1)}(\btheta,t)}{\left\{{S}^{(0)}(\btheta,t)\right\}^2}
\right]^{\rm T},
& \frac{{S}_{{X}}^{(2)}(\btheta,t)}{{S}^{(0)}(\btheta,t)}- \frac{\left\{{S}_{{X}}^{(1)}(\btheta,t)\right\}^2}{\left\{{S}^{(0)}(\btheta,t)\right\}^2}\\
\end{array}
 \right).
\end{eqnarray*}
Therefore, we obtain   that $\sup_{\btheta\in \mathscr{B} , t \in [0 , \tau]}\|\widehat{G}(\btheta,t)-{G}(\btheta,t)\|=o_p(1)$.
Noting that
\begin{eqnarray*}
\frac{\partial \wh{U}({\btheta})}{\partial \btheta}=-\int_{0}^{\tau} \widehat{G}(\btheta,t) {\rm d} \overline{N}(t),
\end{eqnarray*}
we have
\begin{eqnarray*}
\sup_{\btheta\in \mathscr{B} , t \in [0 , \tau]}\left\|\frac{\partial \wh{U}({\btheta})}{\partial \btheta}-\frac{\partial {U}({\btheta})}{\partial \btheta}\right\|=o_p(n).
\end{eqnarray*}
It readily to show that
\begin{eqnarray}\label{A4}
-\frac{1}{n}\frac{\partial \wh{U}({\btheta^*})}{\partial \btheta}\stackrel{p}{\longrightarrow}&\Sigma,
\end{eqnarray}
where $\Sigma=\begin{pmatrix}
  \Sigma_{11} & \Sigma_{12}\\
  \Sigma_{21} & \Sigma_{22}\\
 \end{pmatrix}$ is defined in condition C4.
\\

For the second part of the proof, note that
\begin{eqnarray*}
 \wh{U}(\btheta_0)=\sum_{i=1}^{n}\int_{0}^{\tau}\wh{\bV}_i\,\,{\rm d} N_i(t)
-\int_{0}^{\tau}\frac{\sum_{i=1}^{n} Y_i(t)\wh{\bV}_i\cdot\exp(\wh{\bV}_i^{\rm T}\btheta_0)}{\sum_{i=1}^{n} Y_i(t) \exp(\wh{\bV}_i^{\rm T}\btheta_0)}\,\,{\rm d} \overline{N}(t),
\end{eqnarray*}
and
\begin{eqnarray*}
 {U}(\btheta_0)=\sum_{i=1}^{n}\int_{0}^{\tau}{\bV}_i\,\,{\rm d} N_i(t)
-\int_{0}^{\tau}\frac{\sum_{i=1}^{n} Y_i(t){\bV}_i\cdot\exp({\bV}_i^{\rm T}\btheta_0)}{\sum_{i=1}^{n} Y_i(t) \exp({\bV}_i^{\rm T}\btheta_0)}\,\,{\rm d} \overline{N}(t).
\end{eqnarray*}
Straightforward calculations entail
\begin{eqnarray*}
    & &\wh{U}(\btheta_0) - {U}(\btheta_0)\\
    &=&\left( \begin{array}{c}\mathlarger{\int}_{0}^{\tau}\left\{S_{\bZ}^{(1)}(\btheta_0,t)/S^{(0)}(\btheta_0,t)-\widehat{S}_{\bZ}^{(1)}(\btheta_0,t)/
    \widehat{S}^{(0)}(\btheta_0,t)\right\}{\rm d} \overline{N}(t) \\
      \sum_{i=1}^{n}\mathlarger{\int}_{0}^{\tau}\left\{(\widehat{X}_i-X_i)+S_{X}^{(1)}(\btheta_0,t)/S^{(0)}(\btheta_0,t)-\widehat{S}_{{X}}^{(1)}(\btheta_0,t)/\widehat{S}^{(0)}(\btheta_0,t)
      \right\}{\rm d} N_i(t)\end{array}\right)\\
    &\triangleq & \left( \begin{array}{c}D_1 \\D_2\end{array}\right),
\end{eqnarray*}
where
\begin{eqnarray*}
D_1=\mathlarger{\int}_{0}^{\tau}\left\{S_{\bZ}^{(1)}(\btheta_0,t)/S^{(0)}(\btheta_0,t)-\widehat{S}_{\bZ}^{(1)}(\btheta_0,t)/
    \widehat{S}^{(0)}(\btheta_0,t)\right\}{\rm d} \overline{N}(t),
\end{eqnarray*}
and
\begin{eqnarray*}
D_2=\sum_{i=1}^{n}\mathlarger{\int}_{0}^{\tau}\left\{(\widehat{X}_i-X_i)+S_{X}^{(1)}(\btheta_0,t)/S^{(0)}(\btheta_0,t)-
\widehat{S}_{{X}}^{(1)}(\btheta_0,t)/\widehat{S}^{(0)}(\btheta_0,t)
      \right\}{\rm d} N_i(t).
\end{eqnarray*}
Using the first order expansion yields
\begin{eqnarray*}
\frac{x}{y}&=&\frac{x_0}{y_0} + \frac{x-x_0}{y_0} - \frac{(y-y_0)x_0}{y_0^2} + \textit{o}\l\{ \sqrt{(x-x_0)^2 - (y-y_0)^2}\r\}.
\end{eqnarray*}
Then
\begin{eqnarray*}
    &&\frac{\widehat{S}_{\bZ}^{(1)}(\btheta_0,t)}{\widehat{S}^{(0)}(\btheta_0,t)}-\frac{S_{\bZ}^{(1)}(\btheta_0,t)}{S^{(0)}(\btheta_0,t)}\\
    &=& \frac{\widehat S_{\bZ}^{(1)}(\btheta_0,t) -S_{\bZ}^{(1)}(\btheta_0,t)}{S^{(0)}(\btheta_0,t)}
       - \frac{\l\{\widehat S^{(0)}(\btheta_0,t) - S^{(0)}(\btheta_0,t)\r\}S_{\bZ}^{(1)}(\btheta_0,t)}{\l\{S^{(0)}(\btheta_0,t)\r\}^2}+ \textit{o}_p(1)\\
    &=& \frac{\gamma_0Q_n}{n}\l[\frac{s_{{\bZ}X}^{(1,1)}(\btheta_0,t)}{S^{(0)}(\btheta_0,t)}- \frac{s_X^{(1)}(\btheta_0,t) S_{\bZ}^{(1)}(\btheta_0,t)}
       {\l\{S^{(0)}(\btheta_0,t)\r\}^2}\r] + \textit{o}_p(1)
\end{eqnarray*}
holds uniformly over $t \in[0 , \tau]$.
For simplicity, we define
\begin{eqnarray*}
G_{12}=\frac{S_{{\bZ}X}^{(1,1)}(\btheta_0,t)}{S^{(0)}(\btheta_0,t)}- \frac{S_X^{(1)}(\btheta_0,t) S_{\bZ}^{(1)}(\btheta_0,t)}
          {\l\{S^{(0)}(\btheta_0,t)\r\}^2},
\end{eqnarray*}
then
\begin{eqnarray*}
D_1 &=&\int_{0}^{\tau}-\frac{\gamma_0 Q_n}{n} G_{12} {\rm d} \overline{N}(t)+ \textit{o}_p(1)\\
    &=&-\gamma_0 Q_n \Sigma_{12}+ \textit{o}_p(1).
\end{eqnarray*}
Similarly
\begin{eqnarray*}
D_2 &=&\sum_{i=1}^{n}\int_{0}^{\tau}\left\{(\widehat{X}_i-X_i)+\frac{S_{X}^{(1)}(\btheta_0,t)}{S^{(0)}(\btheta_0,t)}
       -\frac{\widehat{S}_{{X}}^{(1)}(\btheta_0,t)}{\widehat{S}^{(0)}(\btheta_0,t)}\right\}d N_i(t)\\
&=& \text{E}\l\{XN(\tau) \r\}Q_n- \ihe \int_0^{\tau} \frac{S_{X}^{(1)}(\btheta_0,t)}{S^{(0)}(\btheta_0,t)}\cdot \frac{Q_n}{n} \,{\rm d} N_i(t)- \gamma_0
        Q_n\Sigma_{22}+ \textit{o}_p(1)\\
    &=& E \l\{\int_0^{\tau} \l( X - \frac{S_{X}^{(1)}(\btheta_0,t)}{S^{(0)}(\btheta_0,t)} \r) \,{\rm d} N(t) \r\}Q_n- \gamma_0 Q_n\Sigma_{22}+ \textit{o}_p(1)\\
    &=& -\gamma_0 Q_n \Sigma_{22}+ \textit{o}_p(1).
\end{eqnarray*}
Hence,
\begin{eqnarray*}
n^{-1/2}\l\{\wh{U}(\btheta_0) -U(\btheta_0) \r\}
    &=& n^{-1/2}Q_n\binom{-\Sigma_{12} \gamma_0}{-\Sigma_{22} \gamma_0}+ \textit{o}_p(1).
\end{eqnarray*}
Let $ \zeta= (-\Sigma_{12}^{\rm T} \gamma_0, -\Sigma_{22} \gamma_0)^{\rm T}$, then
\begin{eqnarray*}
 n^{-1/2}\wh{U}(\btheta_0)=n^{-1/2} \ihe \int_0^{\tau}\l\{\bV_i - \frac{S^{(1)}(\btheta_0,t)}{S^{(0)}(\btheta_0,t)} \r\} {\rm d}N_i(t)
        +n^{-1/2}Q_n \zeta + \textit{o}_p(1).
\end{eqnarray*}
Set
\begin{eqnarray*}
  \mathscr{F}_t&=&\sigma\l\{N_i(t),N_i^{\rm C}(t),\bZ_i,X_i;\,0\leq t\leq \tau,i=1,\ldots,n\r\}.
\end{eqnarray*}
The processes $M_i(t)$ defined by
\begin{eqnarray*}
  M_i(t)&=& N_i(t)-\int_0^t \lambda_i(u){\rm d}u,\;
  i=1,\ldots,n,\;\; t\in [0,\tau]
\end{eqnarray*}
are local martingales on the time interval $[0,\tau]$. As a consequence, they are in fact local square integrable martingales.
Furthermore, because of $U$ is independent with $(\wt T,C,\bZ,X)$,
we define
\begin{eqnarray*}
U_i(\btheta_0)&=&\int_{0}^{\tau}\l\{\bV_i - \frac{S^{(1)}(\btheta_0,t)}{S^{(0)}(\btheta_0,t)} \r\} \text{d} M_i(t),\; i=1,\ldots,n,
\end{eqnarray*}
then
\begin{eqnarray*}
    & &\text{Cov}\l(n^{-1/2}\,\ihe\, U_i(\btheta_0),n^{-1/2}Q_n\r)\\
    &=& \frac{1}{nE(X)}E\l(\l\{\ihe\,U_i(\btheta_0)\r\} \cdot \l[\ihe \l\{ \wt{X}_i-X_i\r\}\r]\r)\\
    &=&\frac{1}{nE(X)}E \l\{E \l( \l\{\ihe\,U_i(\btheta_0)\r\} \cdot\l[\ihe\,\l\{\wt{X}_i-X_i\r\}\r]\middle|\mathscr{F}_\tau \r) \r\} \\
    &=&\frac{1}{nE(X)}E \l(\l\{\ihe\,U_i(\btheta_0)\r\} \cdot E  \l[\ihe\,\l\{\wt{X}_i-X_i\r\}\middle|\mathscr{F}_\tau \r] \r) \\
    &=&0.
\end{eqnarray*}
Consequently,
\begin{eqnarray*}
\text{Var} \l\{ n^{-1/2}\wh{U}(\btheta_0) \r\}= \text{Var} \l\{n^{-1/2}U(\btheta_0) \r\}+ \text{Var} \l( n^{-1/2} Q_n \zeta \r),
\end{eqnarray*}
where
\begin{eqnarray*}
\text{Var} \l\{n^{-1/2}U(\btheta_0) \r\}\stackrel{p}{\longrightarrow} \Sigma
\end{eqnarray*}
and
\begin{eqnarray*}
\text{Var} \l( \frac{1}{\sqrt n} Q_n \zeta \r)=\frac{\text{Var}(\wt{X}) - \text{Var}(X)}{\l\{E(X)\r\}^2} \zeta \zeta^{\rm T} .
\end{eqnarray*}
As a result, we have
\begin{eqnarray*}
\text{Var} \l\{ \frac{1}{\sqrt n}\wh{ U}(\btheta_0) \r\}
\stackrel{p}{\longrightarrow} \Sigma
+ \frac{\text{Var}(\wt{X}) - \text{Var}(X)}{\l\{E(X)\r\}^2} \zeta \zeta^{\rm T}
\,\triangleq\, \Sigma+\Omega,
\end{eqnarray*}
and
\begin{eqnarray}\label{A5}
\frac{1}{\sqrt n} \wh{ U}(\btheta_0)\stackrel{d}{\longrightarrow} \text{N}(0,\Sigma+\Omega),
\end{eqnarray}
where
\begin{eqnarray*}
\Omega=\frac{\text{Var}(\wt{X}) - \text{Var}(X)}{\l\{E(X)\r\}^2} \zeta \zeta^{\rm T}.
\end{eqnarray*}
Combining (\ref{A4}) and (\ref{A5}), we have
\begin{eqnarray*}
\sqrt{n}(\wh{\btheta}-\btheta_0)&=&\left\{-\frac{1}{n}\frac{\partial \wh{U}(\btheta^*) }{\partial \btheta}\right\}^{-1}\cdot\frac{1}{\sqrt{n}} \wh{U}(\btheta_0)\\
    &\stackrel{d}{\longrightarrow}&\text{N}(0,\Sigma^{-1}(\Sigma+\Omega)\Sigma^{-1}),
\end{eqnarray*}
where
\begin{eqnarray*}
\Sigma^{-1}(\Sigma+\Omega)\Sigma^{-1}&=&\Sigma^{-1}\Sigma\Sigma^{-1}+\frac{\text{Var}(\wt{X})-\text{Var}(X)}{\l\{E(X)\r\}^2}\Sigma^{-1}\zeta \zeta^{\rm T}\Sigma^{-1}\\
    &=&\Sigma^{-1}\Sigma\Sigma^{-1}+\frac{\text{Var}(\wt{X})-\text{Var}(X)}{\l\{E(X)\r\}^2}
         \left(\begin{array}{cccc}
         0&\cdots&0&0\\
         \vdots&&\vdots&0\\
         0&\cdots&0&0\\
         0&\cdots&0&\gamma_0^2\\
         \end{array}\right).
\end{eqnarray*}
We can obtain that
 \begin{eqnarray*}
           \sqrt n (\wh{\bbeta}-{\bbeta}_0)\stackrel{d}{\longrightarrow} \text{N}\l({\bf0},(\Sigma^{-1})_{p}\r),
  \end{eqnarray*}
and
 \begin{eqnarray*}
           \sqrt n (\wh{\gamma}-{\gamma}_0)\stackrel{d}{\longrightarrow} \text{N}\l({0},(\Sigma^{-1})_{(p+1,p+1)}+\frac{\text{Var}(\wt{X})-\text{Var}(X)}{\l\{E(X)\r\}^2}\gamma_0^2\r),
  \end{eqnarray*}
where $(\Sigma^{-1})_{p}$ and $(\Sigma^{-1})_{(p+1,p+1)}$
respectively represent the $p$th order sequential principal minor
and the $(p+1)$th diagonal element of matrix $\Sigma^{-1}$.